# Fixed effects testing in high-dimensional linear mixed models

Jelena Bradic[1] and Gerda Claeskens[2] and Thomas Gueuning[*2]

[1]Department of Mathematics, University of California San Diego
[2]ORStat and Leuven Statistics Research Center, KU Leuven, Belgium

**Abstract**

Many scientific and engineering challenges – ranging from pharmacokinetic drug dosage allocation and personalized medicine to marketing mix (4Ps) recommendations – require an understanding of the unobserved heterogeneity in order to develop the best decision making-processes. In this paper, we develop a hypothesis test and the corresponding p-value for testing for the significance of the homogeneous structure in linear mixed models. A robust matching moment construction is used for creating a test that adapts to the size of the model sparsity. When unobserved heterogeneity at a cluster level is constant, we show that our test is both consistent and unbiased even when the dimension of the model is extremely high. Our theoretical results rely on a new family of adaptive sparse estimators of the fixed effects that do not require consistent estimation of the random effects. Moreover, our inference results do not require consistent model selection. We showcase that moment matching can be extended to nonlinear mixed effects models and to generalized linear mixed effects models. In numerical and real data experiments, we find that the developed method is extremely accurate, that it adapts to the size of the underlying model and is decidedly powerful in the presence of irrelevant covariates.

## 1 Introduction

In many applications, we want to use data to draw inferences about the common underlying effect of a treatment. Examples include medical studies about the computation of effect sizes for assessing the clinical importance of covariates on pharmacokinetic or pharmacodynamic responses, and to the study of drug-drug interactions or studies that quantify the effects of different advertising mediums that take into account other variables such as pricing, distribution points and competitor tactics for example, commonly used by technology firms to optimize budget over these different mediums. Historically, most datasets have been too small to meaningfully explore unobserved heterogeneity beyond dividing the sample into independent subgroups. Recently, however, there has been an explosion of empirical settings where it is potentially feasible to gather large-scale observations and therefore better customize estimates across both population and individuals.

*Support of the Research Foundation Flanders, KU Leuven grant GOA/12/14 and of the IAP Research Network P7/06 of the Belgian Science Policy is acknowledged. Jelena Bradic would like to acknowledge the support of the National Science Foundation grant DMS - 1712481. The computational resources and services used in this work were provided by the VSC (Flemish Supercomputer Center), funded by the Hercules Foundation and the Flemish Government - department EWI.

An impediment to exploring unobserved heterogeneous effects is the fear that researchers will iteratively search for subgroups with high treatment levels, and then report only the results for subgroups with extreme effects, thus exploring and utilizing heterogeneity that may be purely spurious. Moreover, procedural restrictions have often been used to control for the unobserved randomness. However, such procedural restrictions can make it difficult to encompass strong but unexpected heterogeneity which naturally occurs in practice. In this paper, we seek to address this challenge by developing a method for hypothesis testing of fixed effects, while allowing models to have heterogeneous and random components, that yields valid asymptotic confidence intervals and p-values for the true underlying fixed effect.

Classical approaches to test fixed effects in the presence of random effects include Breslow and Clayton (1993), Kenward and Roger (1997) and Crainiceanu and Ruppert (2004); see, e.g., Verbeke and Molenberghs (2009). These methods perform well in applications with a small number of covariates, but quickly break down as the number of covariates increases. In this paper, we explore the use of doubly robust ideas from the literature to improve the performance of these classical methods in the presence of an exploding number of covariates. We develop a family of moment matching tests, which allows for flexible modeling of interactions in high dimensions by allowing models that are not entirely sparse (not even approximately). The developed moments are related to Neyman orthogonalization principles in that they are robust to model misspecifications and estimation of nuisance parameters; however, the tests differ in that they are also robust to the misspecification (and/or mis-estimation) of random effects in the case of linear mixed models – this is especially important in environments with many covariates with possible complex interactions and the presence of unobserved model heterogeneity.

Despite their widespread success in estimation and inference in high-dimensional linear models, there are important hurdles that need to be cleared before penalized estimators are directly useful in linear mixed models. Ideally, an estimator of complete variability in the model needs to be constructed, so that a researcher can use it to test hypothesis and establish sampling distribution. However, in the case of mixed models, estimation of random effects is typically more difficult than that of fixed effects one is interested in. Developing inferential tools that do not necessarily rely on good quality estimates of random effects would therefore be extremely useful. Yet, such procedures have been largely left undeveloped, even in the standard contexts.

This paper addresses these limitations, develops a robust method for fixed effect testing that allows for a tractable asymptotic theory and valid statistical inference even when the estimation of random effects is not consistent. Our proposed test is composed of an estimator of the correlation between the model error under the null hypothesis and the error of carefully constructed feature projections.

In the interest of generality, we begin our theoretical analysis by developing the desired consistency and asymptotic normality results in the context of high-dimensional linear mixed models. We prove these results for a particular variant of mixed models that uses Gaussian random effects with unknown cluster variances while allowing the number of fixed effects to be much larger than the sample size. However, the results do not rely on Gaussianity or quality of estimators of the unknown variance components. This property is achieved by the doubly robust construction of the test statistic, as well as new high-dimensional estimators of the fixed effects. We also show that such robustly motivated estimator is consistent in $l_1$ norm whenever the actual model is sparse. Our proof is built on within cluster analysis and a martingale representation of the test statistic between the clusters. We show that the consistency of the test adapts to the quality of estimation of the fixed effects by successfully leveraging correlation properties among the features. Given these general results, we next show that our ideas extend from the linear setting to the nonlinear setting as well as settings with unknown cluster correlation. We also illustrate that the proposed fixed effect estimator can be utilized for estimation of the unknown variance components.

Although our main focus in this paper is the inference on fixed or common effects, we note that there are a variety of important applications of the p-value construction in a pure variable selection context. Ryzhov et al. (2016) seek to improve the construction of marketing campaigns for non-profit organizations by detecting which marketing strategy yielded a better donor retention (in terms of continuous influx of small to medium donations). Donors are grouped naturally by the observed frequency of donations thus yielding a natural longitudinal structure in the observations. Here we need rigorous predictions for the probability that a donor would continue its activity if a specific marketing strategy was selected as the most beneficial for the observed data. Our results would be one of the first to develop rigorous variable selection that enables the use of large-scale data for this purpose.

## 1.1 Related Work

There has been a longstanding understanding in the high-dimensional statistics literature that prediction methods based on regularization (e.g. Tibshirani (1996) and Fan and Li (2001)) perform well outside of the class of linear models: If the goal is prediction, then we should define a proper likelihood function and the method will be considered as good as long as the likelihood function is convex (see Bühlmann and Van De Geer, 2011, for more details). However, good performance in prediction does not necessarily translate into good performance for estimation or inference about model parameters. In fact, it can often be quite poor. In a Neyman orthogonalization framework we use to formalize our inferential results, we show that the testing of significance of fixed effect can be done asymptotically exactly regardless of the error in estimation of the random effects or the additional nuisance parameters of the model. Thus, when evaluating estimators of the fixed effects, asymptotic theory plays a much more important role than in the standard prediction context.

From a technical point of view, the main contribution of this paper is an asymptotic normality theory enabling statistical inference in the context of high-dimensional linear mixed models while simultaneously allowing misspecification in both the model and the random effects. Recent results by Zhang and Zhang (2014); Van de Geer et al. (2014); Javanmard and Montanari (2014a); Cai et al. (2017); Belloni et al. (2014), and others have established asymptotic properties of tests in a particular variant of the sparse high-dimensional linear models Ren et al. (2015); Jankova et al. (2015); Ning et al. (2017); Athey et al. (2016). To our knowledge, however, we provide the first set of conditions under which tests are both asymptotically unbiased and Gaussian for the linear mixed models, thus allowing for classical statistical inference; the estimator of the fixed effects used to achieve asymptotic normality proposed in this paper is also new. We review the existing theoretical literature on linear mixed models in more detail in Section 3.1.

A small but growing literature, including Belloni et al. (2015); Chernozhukov et al. (2015), has considered the use of Neyman orthogonalization for the purposes of high-dimensional inference. These papers use the Neyman orthogonalization method of Neyman (1959) together with sample splitting and de-biasing (e.g. Chernozhukov et al. (2016)), and report confidence intervals or p-values resulting from significance testing, obtained by Belloni et al. (2017). A limitation of the existing work is in that it cannot allow for the unobserved heterogeneity in the model (like the presence of random effects for example).

We view our contribution as complementary to this literature, by showing that Neyman orthogonalization need not only be viewed as successful in linear models (or generalized linear models), and can instead be modified and used for rigorous asymptotic analysis in high-dimensional model with random effects. We succeed in showing that our construction allows one more degree of robustness – the random effects can be estimated rather poorly without changing the asymptotic null distribution of the test. Even in low dimensions advancements of this type would be of considerable interests as there are no fully reliable ways to identify the best covariance model; as many studies have revealed biased inference for the fixed effects with an underspecified covariance structure; the problem is only multiplied when there is a growing number of fixed effects in the model. The methodological and theoretical tools developed here are useful beyond the specific class of algorithms studied in our paper. In particular, our tools allow for a fairly direct analysis of variants of the Bayesian hierarchical linear regression models (e.g. Quintana et al. (2016)).

Finally, we note a growing literature on estimating fixed effects in the presence of unobserved heterogeneity using different regularization methods. Schelldorfer et al. (2011); Groll and Tutz (2014); Hui et al. (2016) develop lasso-like methods for estimation of fixed effects in a sparse high-dimensional linear mixed model setting. Non-convex methods were investigated by Wang et al. (2012) and Ghosh and Thoresen (2016) among others. Fan and Li (2012) and others discuss variable selection procedures for the fixed or random effects that enable off-the-shelf methods to be used for optimal estimation (see a review article Müller et al. (2013) for more details). Additionally, Bonnet et al. (2015) estimate heritability in sparse linear mixed models, relying on a special asymptotic scenario where the sizes of the clusters are proportional to the number of observations. However, none of the approaches above provides guarantees about honest p-values or confidence interval constructions.

## 2 Inference in Linear Mixed Models

We consider a linear mixed model in which observations are grouped. Suppose that we have a sample of $N$ subjects. Let $i = 1, \dots, N$ be the grouping index and $n_i$ the number of observations in group $i$. The total number of observations is denoted by $n = \sum_{i=1}^{N} n_i$. For the $i$-th subject we represent the model as follows

$$\mathbf{y}_i = \boldsymbol{X}_i \boldsymbol{\gamma}^* + \boldsymbol{Z}_i \beta^* + \mathbf{W}_i \boldsymbol{b}_i + \epsilon_i \tag{1}$$

where $\mathbf{y}_i = (y_{i1}, \dots, y_{in_i})^\top \in \mathbb{R}^{n_i}$ collects the response variable of the $i$-th subject, $\boldsymbol{X}_i \in \mathbb{R}^{n_i \times (p-1)}$ with $\boldsymbol{X}_i = (x_{i1}^\top, x_{i2}^\top, \dots, x_{in_i}^\top)^\top$ the design matrix of the fixed effects where $x_{ij} \in \mathbb{R}^{p-1}$ for $j \in \{1, \dots, n_i\}$. Similarly we denote with $\boldsymbol{Z}_i = (z_{i1}, z_{i2}, \dots, z_{in_i})^\top$ the design vector of the fixed effects which we are interested in, with $z_{ij} \in \mathbb{R}$. The vector of population specific fixed effect coefficients is split into a $(p-1)$-dimensional vector $\boldsymbol{\gamma}^*$ and a univariate $\beta^*$. The subject specific random effects are defined through a $q$ dimensional vector $\boldsymbol{b}_i = (b_{i1}, \dots, b_{iq})^\top \in \mathbb{R}^q$ for which we assume $\boldsymbol{b}_i \sim \mathcal{N}_q(0, \boldsymbol{\psi})$. Here, $\boldsymbol{\psi}$ is the unknown covariance matrix of the random effects which can be correlated with each other. The corresponding deterministic design matrix for group $i$ is denoted with $\boldsymbol{W}_i \in \mathbb{R}^{n_i \times q}$. In addition, $\epsilon_i$ is the random error vector with components independent and identically distributed with mean zero and an unknown variance $0 < \sigma_\epsilon^2 < \infty$. We would like to remark that the independence assumption can be generalized to a positive definite variance covariance structure. This generalization still fits into the theoretical framework presented in Section 3. Nonetheless, for the sake of notational simplicity, we restrict to the case of group independent errors.

The goal of this paper is to develop a test which is able to detect wether $\beta^*$ is equal to $\beta_0$ or not, that is,

$$H_0 : \beta^* = \beta_0 \text{ versus } H_1 : \beta^* \neq \beta_0, \tag{2}$$

for some given value $\beta_0 \in \mathbb{R}$. We study the asymptotics with $N \to \infty$ and assume that $n_i$ is bounded in $N$, while allowing a large number of fixed effects in that $p \gg n$, i.e. $\log p = o(\sqrt{n})$. The main difficulty is that we can only ever test the fixed effects if the random effects are either known or estimated well. However, in practice this is never achievable and we cannot directly test for the fixed effects using existing tools and techniques. In general, we cannot estimate the variance components consistently simply from the observed data without further restrictions on the data generating distribution. A standard way to make progress is to assume models selection consistency of the fixed effects, i.e., that estimated fixed effects are correctly selected. The motivation behind this assumption is that it enables direct dimensionality reduction as it effectively implies correct selection of the true fixed effects; thus, imposing restrictive minimal signal strength assumptions together with a irrepresentable condition. In this paper, we take a more indirect approach: We show that, under simple assumptions, our approach can use moment conditions to achieve consistency of a test without needing to explicitly estimate the variance component.

Let vectors $\boldsymbol{Y}$, $\boldsymbol{b}$ and $\boldsymbol{\epsilon}$, and matrices $\boldsymbol{X}$, $\boldsymbol{Z}$ be obtained by stacking vectors $\mathbf{y}_i$, $\boldsymbol{b}_i$ and $\epsilon_i$ and matrices $\boldsymbol{X}_i, \boldsymbol{Z}_i$ respectively, underneath each other, and let $\boldsymbol{\Psi} = \mathrm{diag}(\boldsymbol{\psi}, ..., \boldsymbol{\psi}) \in \mathbb{R}^{qN \times qN}$ so that $\boldsymbol{W}\boldsymbol{b} \sim \mathcal{N}_n(0, \boldsymbol{W}\boldsymbol{\Psi}\boldsymbol{W}^\top)$ with a block-diagonal matrix $\boldsymbol{W} = \mathrm{diag}(\boldsymbol{W}_1, \ldots, \boldsymbol{W}_N) \in \mathbb{R}^{n \times qN}$. In particular, $\boldsymbol{b} = (\boldsymbol{b}_1^\top, \ldots, \boldsymbol{b}_N^\top)^\top \in \mathbb{R}^{qN}$ is the random effect vector and $\boldsymbol{Y}|(\boldsymbol{X}, \boldsymbol{Z})$ has mean $\boldsymbol{X}\boldsymbol{\gamma}^* + \boldsymbol{Z}\boldsymbol{\beta}^*$ and variance $\sigma_\epsilon^2 \boldsymbol{I}_n + \boldsymbol{W}\boldsymbol{\Psi}\boldsymbol{W}^\top$. We further standardize the design matrix $\boldsymbol{X}$ such that each column has the same norm. The linear mixed effects model (1) can be rewritten as

$$\boldsymbol{Y} = \boldsymbol{X}\boldsymbol{\gamma}^* + \boldsymbol{Z}\beta^* + \boldsymbol{W}\boldsymbol{b} + \boldsymbol{\epsilon}. \tag{3}$$

The $n$ components of the noise vector $\boldsymbol{\epsilon}$ are independent and identically distributed with mean zero and variance $0 < \sigma_\epsilon^2 < \infty$. We assume that $\boldsymbol{\epsilon}$, $\boldsymbol{b}$ and $\boldsymbol{W}$ are mutually independent and that $\boldsymbol{\epsilon}$ is independent of $\boldsymbol{X}, \boldsymbol{Z}$. Observe that we do not require the error to have Gaussian distribution (see Schelldorfer et al., 2011; Fan and Li, 2012, where Gaussianity was explicitly assumed); if we restrict our attention to Gaussian error then the independence above can be replaced with uncorrelatedness with no significant changes in the proofs. Note that $\boldsymbol{\gamma}^*$ and $\beta^*$ can be seen as being derived from an original $p$-dimensional vector. The notation (3) emphasizes the fact that $\beta^*$ is the component of this original vector on which we want to perform hypothesis testing.

## 2.1 From the parametric null to the matching moments condition

At a high level, Wald and score tests can be thought of as two contrasting methods with likelihood based constructions. Given a particular score function (typically a first derivative of the likelihood function), a classical method such as Rao's test, performs estimation only under the null hypothesis and forms a test based on the first moment condition, that is the expectation of the score should be close to zero. In contrast, orthogonalization methods seek to find appropriate directions that are close to the score, where the closeness is defined with respect to a projection of certain kind. The advantage of orthogonalization is that hypothesis testing of a parameter of primary interest can be done rather efficiently in the presence of the nuisance parameters. This is achieved by constructing a class of functions (dependent on the null hypothesis) that is orthogonal to the scores of the nuisance parameters. Neyman then considers a particular construction of regressing the scores of the parameter of interest onto the scores of the nuisance parameters. Then the test, formulated by exploring this orthogonality, can potentially lead a to a substantial increase in power.

In this section, we seek orthogonalization principles that are adapted to the presence of random effects in a linear regression model. Suppose first that we observe independent samples $(\mathbf{y}_i, \boldsymbol{X}_i, \boldsymbol{Z}_i)$ and want to build a orthogonalization criterion suitable for testing (2) in a model (3). We start by splitting the feature space until we have partitioned it into $\boldsymbol{Z}_i$, a set of covariates in the null, and $\boldsymbol{X}_i$, associated with the nuisance parameters. Then, given observations $(\boldsymbol{X}_i, \boldsymbol{Z}_i)$ only, we evaluate the regression of $\boldsymbol{Z}_i$ onto the remaining covariates $\boldsymbol{X}_i$ by setting

$$\boldsymbol{Z} = \boldsymbol{X}\boldsymbol{\theta}^* + \boldsymbol{U} \tag{4}$$

where $\boldsymbol{\theta}^* \in \mathbb{R}^{p-1}$ and with components of $\boldsymbol{U}$ are i.i.d. with mean zero and unknown variance $0 < \sigma_u^2 < \infty$. Heuristically, this strategy is well-motivated if we believe the covariates of the fixed design have a Gaussian distribution and that the rows are roughly identically distributed. There are possibly several other procedures on how to best design this regression; but for simplicity we consider a linear case – other more elaborate cases follow from our methodology with easy extensions of the proofs.

We assume that $\boldsymbol{U}$ is independent of $\boldsymbol{X}$, $\boldsymbol{\epsilon}$ and $\boldsymbol{b}$. However, uncorrelatedness can replace the independence with little changes in the proof whenever the errors in the model take Gaussian structure. Note that the coefficients $\boldsymbol{\theta}^*$ are functions of the variances and covariances of the design matrices $\boldsymbol{X}$ and $\boldsymbol{Z}$, i.e. $\boldsymbol{\theta}^*$ and of $\sigma_u^2$ depend on $\mathcal{X} = (\boldsymbol{X}, \boldsymbol{Z})$. They can be explicitly computed if the rows of $\mathcal{X}$ are i.i.d. with normal distribution $\mathcal{N}_p(0, \boldsymbol{\Sigma})$. In that case, denoting by $j$ the column of $\mathcal{X}$ corresponding to $\boldsymbol{Z}$, it follows from Anderson (1984) that the conditional density $f(Z_i|\boldsymbol{X}_i)$ is a $(p-1)$-variate normal with mean $\boldsymbol{X}_i^\top \boldsymbol{\Sigma}_{-j,-j}^{-1} \boldsymbol{\Sigma}_{-j,j}$ and variance $\sigma_u^2 = \boldsymbol{\Sigma}_{j,j} - \boldsymbol{\Sigma}_{j,-j} \boldsymbol{\Sigma}_{-j,-j}^{-1} \boldsymbol{\Sigma}_{-j,j}$, where $\boldsymbol{\Sigma}_{-j,-j} \in \mathbb{R}^{(p-1)\times(p-1)}$ is obtained by removing the $j$th row and column of $\boldsymbol{\Sigma}$, and $\boldsymbol{\Sigma}_{-j,j} \in \mathbb{R}^{p-1}$ is the $j$th column of $\boldsymbol{\Sigma}$ without its $j$th component. It follows that $\boldsymbol{\theta}^* = \boldsymbol{\Sigma}_{-j,-j}^{-1} \boldsymbol{\Sigma}_{-j,j}$. Table 1 gives the exact values of $\boldsymbol{\theta}^*$ and $\sigma_u^2$ for the setting considered in our numerical work. More details are provided in Appendix.

| Design | $\boldsymbol{\theta}^*$ | Particular case |
|---|---|---|
| Toeplitz | Sparse | $\rho = -0.5 \Rightarrow \begin{cases} \boldsymbol{\theta}^* = (0, \ldots, 0, -0.4, -0.4, 0, \ldots, 0) \\ \sigma_u^2 = 0.6 \end{cases}$ |
| Equi-correlated | Dense | $\rho = 0.8 \Rightarrow \begin{cases} \boldsymbol{\theta}^* = (0.002, \ldots, 0.002) \\ \sigma_u^2 = 0.2 \end{cases}$ |

Table 1: Exact values of $\boldsymbol{\theta}^*$ and $\sigma_u^2$ for the settings considered in our simulation study. Note that for the Toeplitz case, if $\boldsymbol{Z}$ the first (resp. the last) column of $\mathcal{X}$ is tested then $\boldsymbol{\theta}^*$ is slightly different; only its first (resp. last) component is non-zero, with value $\rho$. Furthermore in that case, $\sigma_u^2 = 1 - \rho^2$.

Finally, given regression (3) together with (4) our procedure for testing (2), generates the pseudo response

$$\boldsymbol{V} = \boldsymbol{Y} - \boldsymbol{Z}\beta_0 \tag{5}$$

and observes that under the null hypothesis $\boldsymbol{V} - \boldsymbol{X}\boldsymbol{\gamma}^* = \boldsymbol{W}\boldsymbol{b} + \boldsymbol{\epsilon}$ and $\boldsymbol{Z} - \boldsymbol{X}\boldsymbol{\theta}^* = \boldsymbol{U}$ are uncorrelated and are such that

$$\mathbb{E}[n^{-1}(\boldsymbol{V} - \boldsymbol{X}\boldsymbol{\gamma}^*)^\top \boldsymbol{A}(\boldsymbol{Z} - \boldsymbol{X}\boldsymbol{\theta}^*)] = 0 \tag{6}$$

for a $n \times n$ positive definite matrix $\boldsymbol{A}$. This moment matching equation can be seen as a specific orthogonality condition where $\boldsymbol{Z}_i$ are treated as confounders in the model on $\mathbf{V}$. Conversely, we observe that under the alternative hypothesis the above two terms are correlated since $\boldsymbol{V} - \boldsymbol{X}\boldsymbol{\gamma}^* = \boldsymbol{Z}(\beta^* - \beta_0) + \boldsymbol{W}\boldsymbol{b} + \boldsymbol{\epsilon}$; we compute that

$$\mathbb{E}[n^{-1}(\boldsymbol{V} - \boldsymbol{X}\boldsymbol{\gamma}^*)^\top \boldsymbol{A}(\boldsymbol{Z} - \boldsymbol{X}\boldsymbol{\theta}^*)] = \sigma_u^2 n^{-1} \text{trace}(\boldsymbol{A})(\beta^* - \beta_0).$$

The advantage of the above orthogonalization, (6), is that the form of a moment above resembles double-robust constructions where only one of the two residuals, $\boldsymbol{\epsilon}$ or $\boldsymbol{U}$, needs to be estimated well enough. Moreover, we will see that for a particular choice of the matrix $\boldsymbol{A}$ the above observation leads to optimal inference for the fixed effects without requiring any knowledge or even consistent estimation of the random effects. Such approach is of interest on its own right (even for low-dimensional problems) and would be extremely beneficial for practical purposes in high-dimensional longitudinal studies where estimation of random effects is particularly difficult.

## 2.2 Estimation of the unknowns

In our discussion so far, we have emphasized the flexible nature of our methods: for a wide variety of the structure of $\boldsymbol{\gamma}^*$, distributions of $\boldsymbol{\epsilon}$ and of the random effects, $\boldsymbol{\Psi}$, our methods can be tailored, we achieve both consistency and asymptotic normality, provided the sample size $n$ scales at an appropriate rate. Our results do, however, require the features corresponding to the fixed-effects are not correlated extremely highly: features are sparsely correlated if the precision matrix is row-sparse,

i.e, each row has small number of non-zero elements. We discuss a new class of estimators for both $\boldsymbol{\gamma}^*$ and $\boldsymbol{\theta}^*$ that satisfy this condition and are adaptive to the structure of the random effects.

Our first algorithm, which we call a doubly robust estimator, achieves adaptivity in estimation of $\boldsymbol{\gamma}^*$ regardless of its structure and the consistent estimation of the random effects. It achieves honest estimation by incorporating a pseudo response vector $\boldsymbol{V}$, (5), and a proxy correlation matrix $\widetilde{\boldsymbol{P}}$ directly intro its construction. To motivate our estimation procedure, let us observe that if $\boldsymbol{\epsilon}$ would be normally distributed, the underlying joint density of $(\boldsymbol{V}, \boldsymbol{b})$ would be

$$
\begin{aligned}
f(\boldsymbol{V}, \boldsymbol{b}) &= f(\boldsymbol{V}|\boldsymbol{b}) f(\boldsymbol{b}) \\
&= (2\pi\sigma_\epsilon)^{(-n+qN)/2} |\boldsymbol{\Psi}|^{-1/2} \exp\left\{ -\frac{1}{2\sigma_\epsilon^2} (\boldsymbol{V} - \boldsymbol{X}\boldsymbol{\gamma} - \boldsymbol{W}\boldsymbol{b})^\top (\boldsymbol{V} - \boldsymbol{X}\boldsymbol{\gamma} - \boldsymbol{W}\boldsymbol{b}) - \frac{1}{2} \boldsymbol{b}^\top \boldsymbol{\Psi}^{-1} \boldsymbol{b} \right\}.
\end{aligned} \tag{7}
$$

We treat $f(\boldsymbol{V}, \boldsymbol{b})$ as a quasi-likelihood function; our method does not require normality of $\boldsymbol{\varepsilon}$. For a given $\boldsymbol{\gamma}$, the maximum likelihood estimator of $\boldsymbol{b}$ is then

$$
\widehat{\boldsymbol{b}}(\boldsymbol{\gamma}) = \left( \boldsymbol{W}^\top \boldsymbol{W} + \sigma_\epsilon^2 \boldsymbol{\Psi}^{-1} \right)^{-1} \boldsymbol{W}^\top (\boldsymbol{V} - \boldsymbol{X}\boldsymbol{\gamma}).
$$

We can plug-in this estimator into (7) to construct a profile likelihood for $\boldsymbol{\gamma}$. We define $\boldsymbol{E} = \boldsymbol{W}^\top \boldsymbol{W} + \sigma_\epsilon^2 \boldsymbol{\Psi}^{-1}$ and $\boldsymbol{P} = (\boldsymbol{I}_n - \boldsymbol{W}\boldsymbol{E}^{-1}\boldsymbol{W}^\top)(\boldsymbol{I}_n - \boldsymbol{W}\boldsymbol{E}^{-1}\boldsymbol{W}^\top) + \sigma_\epsilon^2 \boldsymbol{W}\boldsymbol{E}^{-1}\boldsymbol{\Psi}^{-1}\boldsymbol{E}^{-1}\boldsymbol{W}^\top$ and, dropping the constants, we obtain the following profile log-likelihood:

$$
\mathcal{L}(\boldsymbol{\gamma}, \widehat{\boldsymbol{b}}(\boldsymbol{\gamma})) = -\frac{1}{2\sigma_\epsilon^2} (\boldsymbol{V} - \boldsymbol{X}\boldsymbol{\gamma})^\top \boldsymbol{P} (\boldsymbol{V} - \boldsymbol{X}\boldsymbol{\gamma}). \tag{8}
$$

By Lemma 1, $\boldsymbol{P} = (\boldsymbol{I}_n + \sigma_\epsilon^{-2} \boldsymbol{W}\boldsymbol{\Psi}\boldsymbol{W}^\top)^{-1}$. Note that if $\boldsymbol{\epsilon}$ is normally distributed then $\boldsymbol{W}\boldsymbol{b} + \boldsymbol{\epsilon} \sim \mathcal{N}_n(0, \sigma_\epsilon^2 \boldsymbol{P}^{-1})$. However, observe that the above profile log-likelihood is a non-convex function of all of the unknown parameters, $\boldsymbol{\gamma}, \boldsymbol{\Psi}, \sigma_\epsilon$. Moreover, in the presence of high-dimensional fixed effects, the profile log-likelihood has too many stationary points.

Following the idea of Fan and Li (2012), we replace the unknown variance matrix $\sigma_\epsilon^{-2}\boldsymbol{\Psi}$ by a known fixed symmetric matrix $\boldsymbol{M}$ which serves as a proxy. We discuss in the next section how to choose $\boldsymbol{M}$; our theory allows for many choices of $\boldsymbol{M}$. We define proxy versions of $\boldsymbol{E}$, $\boldsymbol{P}$ and $\mathcal{L}$ by replacing $\sigma_\epsilon^{-2}\boldsymbol{\Psi}$ by $\boldsymbol{M}$; we define $\widetilde{\boldsymbol{E}} = \boldsymbol{W}^\top \boldsymbol{W} + \boldsymbol{M}^{-1}$ and $\widetilde{\boldsymbol{P}} = (\boldsymbol{I}_n + \boldsymbol{W}\boldsymbol{M}\boldsymbol{W}^\top)^{-1}$ and obtain the following proxy log-likelihood:

$$
\widetilde{\mathcal{L}}(\boldsymbol{\gamma}, \widehat{\boldsymbol{b}}(\boldsymbol{\gamma})) = -\frac{1}{2\sigma_\epsilon^2} (\boldsymbol{V} - \boldsymbol{X}\boldsymbol{\gamma})^\top \widetilde{\boldsymbol{P}} (\boldsymbol{V} - \boldsymbol{X}\boldsymbol{\gamma}). \tag{9}
$$

Observe that the proxy log-likelihood does not have a good quality approximation of the true $\mathcal{L}$. Instead it serves as a proxy to give inspiration for a new estimator of the unknown fixed effects $\boldsymbol{\gamma}^*$. Provided that the $l_1$ penalization of the above proxy log-likelihood heavily depends on the correct model specification and the form of the likelihood function, we take on a different perspective and design a new estimator that is inspired by the Dantzig selector. We define $\widehat{\boldsymbol{\gamma}}$ as the solution to the following optimization problem

$$
\begin{aligned}
\widehat{\boldsymbol{\gamma}} = \quad & \operatorname{argmin}_{\boldsymbol{\gamma} \in \mathbb{R}^{p-1}} && \|\boldsymbol{\gamma}\|_1 \\
& \text{subject to} && \left\| n^{-1} \boldsymbol{X}^\top \widetilde{\boldsymbol{P}} (\boldsymbol{V} - \boldsymbol{X}\boldsymbol{\gamma}) \right\|_\infty \le \eta_\gamma \\
& && n^{-1} \boldsymbol{V}^\top \widetilde{\boldsymbol{P}} (\boldsymbol{V} - \boldsymbol{X}\boldsymbol{\gamma}) \ge \bar{\eta}_\gamma \\
& && \left\| \widetilde{\boldsymbol{P}} (\boldsymbol{V} - \boldsymbol{X}\boldsymbol{\gamma}) \right\|_\infty \le \mu_\gamma
\end{aligned} \tag{10}
$$

with $\eta_\gamma \asymp (\log n)\sqrt{n^{-1}\log p}$, $\mu_\gamma \asymp \sqrt{\log n}$ and $0 < \bar{\eta}_\gamma < n^{-1}\text{trace}(\sigma_\epsilon^2 \widetilde{\boldsymbol{P}}\boldsymbol{P}^{-1})$ suitably chosen tuning parameters. The $l_1$ norm induces sparse solutions whereas the constraints ensure good theoretical properties. The above constraints arise from the score vector of the profile-log likelihood function $\widetilde{\mathcal{L}}$; the first ensures that the score vector is minimized and the second ensures that the variance of the residuals is properly guessed. However, observe that we do not assume that the profile log likelihood is correctly specified, and for that matter a last constraint is needed to guarantee small size of the estimated residuals in the model

$$\widetilde{\boldsymbol{P}}^{1/2}\boldsymbol{V} = \widetilde{\boldsymbol{P}}^{1/2}\boldsymbol{X}\boldsymbol{\gamma} + \boldsymbol{e}, \tag{11}$$

where $\text{var}(\boldsymbol{e}) = \sigma_\epsilon^2 \widetilde{\boldsymbol{P}}\boldsymbol{P}^{-1}$. The tuning parameter $\bar{\eta}_\gamma$ provides a proxy for the unknown variance of the error in the model above. Hence, the size of the residuals is constrained in a similar way as robust loss functions are truncated to down-weight the effect of overly large outliers.

**Remark 1.** *The estimator $\widehat{\boldsymbol{\gamma}}$ defined in (10) is new and of potential interest in its own right. We show that under suitable conditions, $\|\widehat{\boldsymbol{\gamma}} - \boldsymbol{\gamma}^*\|_1 = O_P(\eta_\gamma \|\boldsymbol{\gamma}^*\|_0)$ meaning that this estimator is consistent if $\|\boldsymbol{\gamma}^*\|_0 = o(\sqrt{n}/\log(p)/\log(n))$. It is worth pointing that this result does not impose any restrictions on the minimum signal strength or that the matrix $\widetilde{\boldsymbol{P}}$ correctly specifies variance parameters of the random effects and in that sense is honest and robust. Beside providing a reliable estimator, the optimization problem (10) is fast since it can be reformulated as a linear program. Existing regularized schemes, such as the one introduced by Schelldorfer et al. (2011) for example, do not provide such quick implementation.*

The estimation of $\boldsymbol{\theta}^*$ is based on model (4). Unlike linear models, where node-wise lasso is sufficient to estimate feature correlation, for the case of linear mixed models, we observe that the matrix $\widetilde{\boldsymbol{P}}$ should contribute to the estimation procedure – we see from (11) that the covariates are premultiplied with $\widetilde{\boldsymbol{P}}^{1/2}$. Namely, the presence of random effects induces a larger than usual dependence in the design matrix of the fixed effects. We incorporate such requirement in a new estimator defined below.

$$\begin{aligned} \widehat{\boldsymbol{\theta}} = \quad & \text{argmin}_{\theta \in \mathbb{R}^{p-1}} & \|\boldsymbol{\theta}\|_1 \\ & \text{subject to} & \left\|n^{-1}\boldsymbol{X}^\top(\boldsymbol{Z} - \boldsymbol{X}\boldsymbol{\theta})\right\|_\infty \le \eta_\theta \\ & & \left\|n^{-1}\boldsymbol{X}^\top\widetilde{\boldsymbol{P}}(\boldsymbol{Z} - \boldsymbol{X}\boldsymbol{\theta})\right\|_\infty \le \eta'_\theta \\ & & n^{-1}\boldsymbol{Z}^\top(\boldsymbol{Z} - \boldsymbol{X}\boldsymbol{\theta}) \ge \bar{\eta}_\theta \\ & & \|\boldsymbol{Z} - \boldsymbol{X}\boldsymbol{\theta}\|_\infty \le \mu_\theta \end{aligned} \tag{12}$$

with tuning parameters $\eta_\theta, \eta'_\theta \asymp (\log n)\sqrt{n^{-1}\log p}$, $\mu_\theta \asymp \sqrt{\log n}$ and $0 < \bar{\eta}_\theta < \sigma_u^2$.

The first and the second constraint enable adaptive and automatic estimation of the correlation of the features of the fixed effects without prior knowledge (or correct estimation) of the correlation of the random effects. In particular, the first ensures that the gradient of the square loss is close to zero, whereas the second one looks at the reweighed gradient where the weights are defined through the matrix $\widetilde{\boldsymbol{P}}$. The presence of the second constraint allows us additional flexibility with the choice of the matrix $\widetilde{\boldsymbol{P}}$. We can remove the second constraint if we impose that $\lambda_{\max}(\boldsymbol{W}^\top\boldsymbol{M}\boldsymbol{W})$ is bounded; however, this would restrict our choices of the matrix $\boldsymbol{M}$ and in particular it would not allow for $\boldsymbol{M} = \log(n)\mathbb{I}_n$ which we found to be beneficial whenever the variance in the random effects is particularly large. Lastly, the last constraint excludes features $\boldsymbol{X}$ that are highly correlated with $\boldsymbol{Z}$; as our task was to remove heterogeneity (by de-correlating features), such reasoning is needed for a successful test statistic.

## 2.3 Asymptotic inference for linear mixed models

Our results are achieved under the asymptotics where $N \to \infty$ and $n_i/N \to 0$, a regime different from that used in simple linear models, and require very mild conditions on the choice of the matrix $\widetilde{\boldsymbol{P}}$, i.e., the matrix $\boldsymbol{M}$, which needs to have the same diagonal structure as $\boldsymbol{W}\boldsymbol{W}^\top$. However, given these high level conditions, we obtain a widely applicable result that applies to several different linear mixed models.

In order to define a test statistic, we focus on the moment condition (6) and in particular consider $\mathbf{A} = \widetilde{\boldsymbol{P}}$. That is, the moment condition that we wish to test is now taking the form $\mathbb{E}[n^{-1}(\boldsymbol{V} - \boldsymbol{X}\boldsymbol{\gamma}^*)^\top \widetilde{\boldsymbol{P}}(\boldsymbol{Z} - \boldsymbol{X}\boldsymbol{\theta}^*)] = 0$ whenever the null hypothesis $H_0$ hold. For the alternative hypothesis, this moment takes the form $\mathbb{E}[n^{-1}(\boldsymbol{V} - \boldsymbol{X}\boldsymbol{\gamma}^*)^\top \widetilde{\boldsymbol{P}}(\boldsymbol{Z} - \boldsymbol{X}\boldsymbol{\theta}^*)] = \sigma_u^2 n^{-1}\text{trace}(\widetilde{\boldsymbol{P}})(\beta^* - \beta_0)$.

We then proceed to form a test statistic. The moment above produces a doubly robust test statistic as long as one of two component regression models is correctly specified and assuming that there are no unmeasured confounders, giving the analyst two chances to correctly specify at least one of the regression models, (3) and (4), respectively. In order to standardize the test statistic appropriately, we define $\widehat{\sigma}^2 = n^{-1}\|\widetilde{\boldsymbol{P}}(\boldsymbol{V} - \boldsymbol{X}\widehat{\boldsymbol{\gamma}})\|_2^2$ and $\widehat{\sigma}_u^2 = n^{-1}\|(\boldsymbol{Z} - \boldsymbol{X}\widehat{\boldsymbol{\theta}})\|_2^2$. Whenever the models are correctly specified, they produce residual variance estimates. However, the quality of testing does not rely on them being consistent estimators; in fact, as the proxy log-likelihood is not exact, this will not be achievable; we show in Lemma 7 that $\widehat{\sigma}^2 = \sigma_\epsilon^2 n^{-1}\text{trace}(\widetilde{\boldsymbol{P}}\boldsymbol{P}^{-1}\widetilde{\boldsymbol{P}}) + o_P(1)$.

Now, we proceed to define the test statistic as

$$T_n = \frac{n^{-1/2}(\boldsymbol{Z} - \boldsymbol{X}\widehat{\boldsymbol{\theta}})^\top \widetilde{\boldsymbol{P}}(\boldsymbol{V} - \boldsymbol{X}\widehat{\boldsymbol{\gamma}})}{\widehat{\sigma}_u \widehat{\sigma}}. \tag{13}$$

The doubly robust test statistic combines the initial model under the null hypothesis with a regression model of the relationship between covariates and the feature related to each of the parameters of interest in such a way that, as long as either the initial model or the feature regression model is correctly specified, the effect of the initial null hypothesis on the transformed moment is correctly estimated. In the case of linear mixed models, the initial model is only correctly specified when the variance parameters $\boldsymbol{\Psi}$ are known, making the test statistic above particular useful for cases when there is no such knowledge.

Our first result is that the test statistic has asymptotically a standard normal distribution whenever the null hypothesis holds (see Theorem 3). For this result we do not require sparsity of the fixed effects, but rather sparsity of the precision matrix of the design matrix of the fixed effects. However, we do not require any consistent estimation of the covariance parameters $\boldsymbol{\Psi}$. To obtain the asymptotic distribution under the alternative hypothesis, $H_1 : \beta^* = \beta_0 + n^{-1/2}h$, we need to assume that the vector of nuisance parameters $\boldsymbol{\gamma}^*$ is sparse (or approximately) with a small number of non-zero elements, although our simulations show that the power is preserved even if the fixed effects have as many as $n$ non-zero elements (see Theorem 4). To our knowledge, all existing results regarding inference of linear mixed models require consistent estimation of the covariance parameters and low-dimensionality of the fixed effects vector. Although some empirical work has demonstrated that inconsistent variance estimation does not effect type I error control (in low-dimensional settings), theoretical guarantees were never established.

## 3 Asymptotic Theory for Linear Mixed Models

In order to use the test statistics to provide formally valid statistical inference, we need an asymptotic normality theory in the desired asymptotics. We begin by precisely describing the asymptotics under the null hypothesis. We then proceed to describe the asymptotics under the alternative hypothesis

$$H_1 : \beta^* = \beta_0 + hn^{-1/2}. \tag{14}$$

Before stating our theoretical results, we give some definitions.

**Definition 1.** *We say that a symmetric semi-definite positive matrix* $\boldsymbol{A}$ *satisfies the* $P$-condition *if it has the same block diagonal structure as* $\boldsymbol{W}^\top \boldsymbol{W}$ *and if each element of* $\boldsymbol{A}$ *is* $\mathcal{O}(\log(n))$.

In Lemma 4, we show that $\widetilde{\boldsymbol{P}} = (\boldsymbol{I}_n + \boldsymbol{W}\boldsymbol{M}\boldsymbol{W}^\top)^{-1}$ satisfies the $P$-condition for any matrix $\boldsymbol{M}$ having the same block diagonal structure as $\boldsymbol{W}^\top \boldsymbol{W}$. Furthermore, it holds that $\boldsymbol{P}$ and $\boldsymbol{P}^{-1}$ satisfy the $P$-condition.

Next, we require that the errors $\boldsymbol{\epsilon}$ in our model have sub-Gaussian tails. A test statistic, as defined in Section 2.3, can be used to give honest p-values or confidence intervals; in Section 4 we illustrate its good properties even when the error is not sub-Gaussian but exhibits heavy-tails.

**Definition 2.** *A random variable* $X$ *is said to have an exponential-type tail with parameters* $(b, \gamma)$ *if* $\forall x > 0$, $\mathbb{P}(|X| > x) \leq \exp[1 - (x/b)^\gamma)]$. *Furthermore, a random variable* $X$ *is sub-Gaussian if it has an exponential-type tail with parameters* $(b, 2)$.

In order to guarantee consistency, we also need to enforce that the design matrix of the fixed effects satisfies some regularity conditions. The conditions are standard in the high-dimensional literature and ensure that the population feature covariance matrix is a well conditioned matrix. Here, we follow Rudelson and Zhou (2013), and achieve this effect by enforcing sub-Gaussianity in the design in the following condition. Let $C_{\min}$, $C_{\max}$ and $\kappa$ denote positive constants.

**Condition 1.** *The matrix* $\boldsymbol{\Sigma} = \mathbb{E}[\boldsymbol{X}^\top \boldsymbol{X}]/n \in \mathbb{R}^{(p-1)\times(p-1)}$ *is such that* $C_{min} \leq \sigma_{min}(\boldsymbol{\Sigma}) \leq \sigma_{max}(\boldsymbol{\Sigma}) \leq C_{max}$ *with* $\sigma_{min}(\boldsymbol{\Sigma})$ *and* $\sigma_{max}(\boldsymbol{\Sigma})$ *the minimal and maximal singular values of* $\boldsymbol{\Sigma}$. *The vectors* $\boldsymbol{\Sigma}^{-1/2}x_i$, $\boldsymbol{\epsilon}$ *and* $\boldsymbol{u}$ *are centered with sub-Gaussian norms upper bounded by* $\kappa$. *Moreover,* $\sigma_u, \sigma_\epsilon \in [C_{min}, C_{max}]$. *The elements of the symmetric and invertible matrix* $\boldsymbol{\Psi}$ *and of the deterministic design matrix* $\boldsymbol{W}$ *are bounded.*

The remaining definition is more technical. We use a regularity condition to control the shape of the correlation in the features and in the variance covariance matrix $\boldsymbol{\Psi}$. Such condition is used regularly in high-dimensional estimation and inference (see Bickel et al., 2009).

**Definition 3.** *We say that the restricted eigenvalue condition holds for a triplet* $(s, \kappa, \boldsymbol{X})$ *if*

$$\min_{\substack{J_0 \subseteq \{1,\ldots,p-1\} \\ |J_0| \leq s}} \quad \min_{\substack{\boldsymbol{\delta} \neq 0 \\ \left\|\boldsymbol{\delta}_{J_0^c}\right\|_1 \leq \left\|\boldsymbol{\delta}_{J_0}\right\|_1}} \frac{\|\boldsymbol{X}\boldsymbol{\delta}\|_2}{\sqrt{n}\,\|\boldsymbol{\delta}_{J_0}\|_2} \geq \kappa.$$

Rudelson and Zhou (2013) show that the restricted eigenvalue condition holds with large probability if the sample size $n$ is large enough and the rows of the design have sub-Gaussian distributions.

We note that our test statistic (13) does not depend on the inverse of $\boldsymbol{\Psi}$. So although we started this section with the nonsingularity assumption of $\boldsymbol{\Psi}$, in practice our method can be directly applied even when noise random effects exist.

Given these preliminaries, we state our main result on the asymptotic normality of the test statistic $T_n$. As discussed previously, we require that the sample size scales as $\sqrt{n}/\log(p) = o(1)$ as $N \to \infty$. If the subsample size grows slower than this, the test statistic will still be asymptotically normal, but the it may be asymptotically biased. Although we treat $n, s$ and $p$ as functions of $N$, for clarity, we state the following result without this explicit dependence.

**Theorem 1.** *Suppose that we have* $N$ *independent observations* $(\mathbf{y}_i, \boldsymbol{X}_i, \boldsymbol{Z}_i, \mathbf{W}_i) \in \mathbb{R}^{n_i} \times \mathbb{R}^{n_i \times (p-1)} \times \mathbb{R}^{n_i} \times \mathbb{R}^{n_1 \times q}$. *Suppose moreover that the features satisfy Condition 1. Let the matrix* $\widetilde{\boldsymbol{P}}$ *be such that Definition 1 holds. Given this data-generating process, let* $\widehat{\gamma}$ *and* $\widehat{\boldsymbol{\theta}}$ *be estimators of the nuisance parameters and the feature dependence as defined in* (10) *and* (12). *Finally, suppose that the sample* *size scales as* $n^{1/2}/\log(p)/\log(n) = o(1)$ *and* $\max_i n_i/N = o(1)$ *and that the models are normalized so that* $\|\boldsymbol{\gamma}^*\|_2 = O(1)$ *and* $\|\boldsymbol{\theta}^*\|_2 = O(1)$*. Then,*

$$T_n \stackrel{d}{=} \mathcal{N}(0,1) + \sqrt{n}(\beta^* - \beta_0)\frac{n^{-1}\text{trace}(\widetilde{\boldsymbol{P}})}{\sqrt{n^{-1}\text{trace}(\widetilde{\boldsymbol{P}}\boldsymbol{P}^{-1}\widetilde{\boldsymbol{P}})}} + o_P(r_n),$$

*where* $r_n = n^{1/2}\eta_\theta\eta_\gamma \|\boldsymbol{\theta}^*\|_0 \kappa^{-2}\bar{\eta}_\gamma^{-1}\sigma_\epsilon/\sigma_u$ *whenever* $\beta^* = \beta_0$ *and* $r_n = n^{1/2}\eta_\gamma \|\boldsymbol{\gamma}^*\|_0 \kappa^{-2}\eta'_\theta\sigma_\epsilon/\sigma_u + \eta_\theta\sqrt{\|\boldsymbol{\theta}^*\|_0}\sigma_\epsilon/\sigma_u + \sqrt{n}(\beta^* - \beta_0)N/n\sigma_\epsilon/\sigma_u$ *whenever* $\beta^* \neq \beta_0$*.*

The construction of p-values is based on the asymptotic distribution of the test statistic $T_n$. For the null hypothesis $H_0 : \beta^* = \beta_0$ , we define the p-value for the two-sided alternative as $P_0 = 2(1 - \Phi(|T_n(\beta_0|)))$. Of course, we could also consider one-sided alternatives with an obvious modification. Whenever the conditions of the Theorem 1 hold, then for any $0 < \alpha < 1$,

$$\limsup_{n\to\infty} P\left[P_0 \le \alpha\right] = \alpha \text{ if } H_0 \text{ holds .}$$

Furthermore, for any sequence $a_n \to 0$ ( $n \to \infty$) which converges sufficiently slowly, the statements also hold when replacing $\alpha$ by $a_n$. A discussion about detection power of the method is given in Section 3.

The proof of Theorem 1 is organized as follows. In Section 3.2, we provide bounds for the $l_1$ norm error in estimation of both $\boldsymbol{\gamma}^*$ and $\boldsymbol{\theta}^*$, while Section 3.3.1 studies the sampling distributions of the test statistic under the null hypothesis and establishes asymptotic Gaussianity. Given a subsampling rate and sparsity requirements in the two models, (3) and (4), we showcase the asymptotic distribution under the alternative hypothesis in Section 3.3.2. Before beginning the proof, however, we relate our result to existing results about linear mixed models in Section 3.1.

## 3.1 Theoretical background

There has been considerable work in understanding the theoretical properties of linear mixed models. The convergence and consistency properties of estimation have been studied by, among others, McCulloch (1997); McGilchrist (1994); Lindstrom and Bates (1988) in low-dimensional setting, Goeman et al. (2011) with growing dimensions and Schelldorfer et al. (2011) in high-dimensional setting. Meanwhile, the inference has been analyzed by Breslow and Clayton (1993) in low-dimensional setting. However, to our knowledge, our Theorem 1 is the first result establishing conditions under which inference in linear mixed models can be performed despite the presence of a large number of fixed effects and taking into account model selection. Moreover, Theorem 1 establishes this result without even resorting to a proper (let alone consistent) estimation of the random effects therefore enabling inference in linear models with misspecified random effects (the structure or distribution); we believe that such result is unique in inferential theory of linear mixed models.

Probably the closest existing result is that of Fan and Li (2012), who showed that selection of fixed effects can be achieved without resorting to proper estimation of random effects. They establish model selection consistency results under conditions similar to ours; however, we note that they impose a slightly more restrictive choice of the matrix $\widetilde{\boldsymbol{P}}$ whereas we allow more general choices – for example choice of $\boldsymbol{M}$ as an identity matrix is not allowed in their work, but it is in ours. Moreover, the authors therein do not discuss asymptotic distributions and hypothesis testing problems. Thus, their results cannot be used for valid asymptotic statistical inference about fixed effects.

Inference for linear mixed models that is robust to the normality assumption in random effects or the model error has been studied exclusively in low-dimensional setting in Zhang and Davidian (2001), who showed that normality can be substituted with a broader class of distributions and an EM algorithm can be used for estimation. However, their results still heavily depend on a correct likelihood specification. Identical conclusions hold for a class of semi-parametric linear mixed models introduced in Lachos et al. (2010) or a linear mixed model with random mixture of normals (Verbeke and Lesaffre, 1996; Song et al., 2007). See Heagerty and Kurland (2001) for an illustration of the effects of inconsistent variance estimation for the likelihood based inference on fixed effects. Additional work on this topic Litière et al. (2007), Hobert and Casella (1996) had only highlighted connections to statistical tests and the effects of misspecification of the random effects, but did not establish any formal justification for it.

### 3.2 Estimation properties

We start by bounding the bias of the newly introduced high-dimensional estimators. Our approach relies on showing that the true parameter vector lies in the constraint set of the problems (10) and (12) for appropriate choices of the tuning parameters defined therein. A sparsity assumption then allows us to bound the bias. We show that the error of estimation is optimal and is not effected by the misspecification of the random effects. In order to state the result, define $\|\boldsymbol{\gamma}^*\|_0$ as the number of non-zero elements of a vector $\boldsymbol{\gamma}^*$.

**Theorem 2.** *Suppose $\widetilde{\boldsymbol{P}} \in \mathbb{R}^{n\times n}$ satisfies the $P$-condition. Suppose that Condition 1 holds and that there exists $\kappa > 0$ such that the restricted eigenvalue condition holds for $(\|\boldsymbol{\gamma}^*\|_0, \kappa, \widetilde{\boldsymbol{P}}^{1/2}\boldsymbol{X})$. Then there exist constants $C_1$, $C_2$, $c_0 > 0$ such that for any $\eta_\gamma \geq C_1 \log(n)\sqrt{n^{-1}\log p}$, $\mu_\gamma \geq C_2\sqrt{\log n}$ and $\bar{\eta}_\gamma, \in (0, n^{-1}\text{trace}(\sigma_\epsilon^2 \widetilde{\boldsymbol{P}}\boldsymbol{P}^{-1}) - c_0)$, for a large enough constant $C_3$ it holds with probability $1 - p^{-C_3}$*

$$\|\widehat{\boldsymbol{\gamma}} - \boldsymbol{\gamma}^*\|_1 \leq 8\eta_\gamma \|\boldsymbol{\gamma}^*\|_0 \kappa^{-2}, \text{ and} \tag{15}$$

$$n^{-1}\left\|\widetilde{\boldsymbol{P}}^{1/2}\boldsymbol{X}(\widehat{\boldsymbol{\gamma}} - \boldsymbol{\gamma}^*)\right\|_2^2 \leq 16\eta_\gamma^2 \|\boldsymbol{\gamma}^*\|_0 \kappa^{-2} \quad . \tag{16}$$

This theorem then directly translates into a bound on the bias in estimation. Namely, with $\eta_\gamma \asymp \log(n)\sqrt{n^{-1}\log p}$, $\mu_\gamma \asymp \sqrt{\log n}$ and $0 < \bar{\eta}_\gamma < n^{-1}\text{trace}(\sigma_\epsilon^2 \widetilde{\boldsymbol{P}}\boldsymbol{P}^{-1})$ the estimator $\widehat{\boldsymbol{\gamma}}$ of a $s$ sparse vector $\boldsymbol{\gamma}^*$,

$$\|\widehat{\boldsymbol{\gamma}} - \boldsymbol{\gamma}^*\|_1 = \mathcal{O}_P(s\log(n)\sqrt{\log p/n}).$$

The conditions required for the above result seem very mild. Finite sample oracle risk properties have been established in Schelldorfer et al. (2011); however, the result therein crucially depends on correct estimation of variance components. In contrast, our results hold for a wide variety of choices of variance estimators; in particular, it remains valid for even non-consistent estimators – the only assumption is on the structure of the matrix $\mathbf{P}$ illustrated through the P-condition. Furthermore, the rate above matches the rate obtained by the mle estimator of Schelldorfer et al. (2011) highlighting excellent robustness properties of the proposed estimator $\widehat{\boldsymbol{\gamma}}$ and indicating optimality at esitmation.

### 3.3 Testing properties

Our analysis proceeds in two steps: the first discusses asymptotic normality of the test statistic whenever the null hypothesis holds, whereas the second discusses the case under the alternative hypothesis.

#### 3.3.1 Size

Analyzing specific linear mixed models can be challenging especially if the model (1) is not fully or correctly specified. Below we establish asymptotic normality of the proposed test statistic. The result is independent of the size of the sparsity of the nuisance parameters and in particular holds for fully dense and ultra high-dimensional vectors. Moreover, it does not require consistent estimation of the variance components.

**Theorem 3.** *Suppose* $\widetilde{\boldsymbol{P}} \in \mathbb{R}^{n\times n}$ *satisfies the* $P$*-condition. Suppose that Condition 1 holds and that there exists* $\kappa > 0$ *such that the restricted eigenvalue condition holds for* $(\|\boldsymbol{\theta}^*\|_0, \kappa, \widetilde{\boldsymbol{P}}^{1/2}\boldsymbol{X})$*. Furthermore, assume that* $\|\boldsymbol{\theta}^*\|_0 = o(\sqrt{n}/\log(p)/\log(n))$ *and a normalization of the form* $\|\boldsymbol{\gamma}^*\|_2 = \mathcal{O}(1)$ *and* $\|\boldsymbol{\theta}^*\|_2 = \mathcal{O}(1)$*. Then, under the null hypothesis* $H_0 : \beta_* = \beta_0$*, it holds that* $T_n \xrightarrow{d} \mathcal{N}(0, 1)$*.*

Theorem 3 establishes the asymptotic distribution of our test statistics under the null hypothesis; it does not require minimal signal strength in the model (1) or an irrepresentable condition. Notably, our statistic does not require $\boldsymbol{\gamma}^*$ to be sparse. This is a strong and unusual result in high-dimensions. This finding is further illustrated numerically in Section 4 where we observe stable control over the type I error rate regardless of the size of $\|\boldsymbol{\gamma}^*\|_0$.

Furthermore, using some arguments of the proof of Theorem 4, interchangeability of sparsity conditions can be shown: if $\boldsymbol{\gamma}^*$ is sparse then $\boldsymbol{\theta}^*$ does not need to be sparse. Note that the $o(\sqrt{n}/\log(p)/\log(n))$ sparsity rate is up to a $\log(n)$ term matching those of simple linear models, see for instance van de Geer et al. (2014) and Javanmard and Montanari (2014b). The additional $\log(n)$ term is needed in controlling the random effects and can be though of as a price to pay for being able to provide optimal estimation despite the correct specification of the random effects.

Besides providing the size property of the $T_n$ statistic, Theorem 3 can also be used to construct confidence intervals. The $1 - \alpha$ confidence interval for $\beta^*$ can be defined as

$$\left\{\beta : 1 - \Phi^{-1}(1 - \alpha/2) \leq T_n(\beta) \leq \Phi^{-1}(1 - \alpha/2)\right\}$$

where $\Phi$ is the standard Gaussian cumulative distribution function and where $T_n(\beta)$ is the test statistic derived under the null hypothesis $\beta^* = \beta$.

#### 3.3.2 Power

In the previous section, we showed that the type I error is asymptotically equal to $\alpha$ for a given level $\alpha \in (0, 1)$. In this section, we show that the test statistic furthermore has tight control of the type II error and preserves power asymptotically while allowing inconsistent variance estimation. In this regard we believe that our result stands out in the existing literature. In addition, we quantify the asymptotic efficiency loss due to misspecification of the random effects.

**Theorem 4.** *Suppose* $\widetilde{\boldsymbol{P}} \in \mathbb{R}^{n\times n}$ *satisfies the* $P$*-condition, Condition 1 holds and that there exists* $\kappa > 0$ *such that the restricted eigenvalue condition holds for* $(\|\boldsymbol{\theta}^*\|_0 \vee \|\boldsymbol{\gamma}^*\|_0, \kappa, \widetilde{\boldsymbol{P}}^{1/2}\boldsymbol{X})$*. Furthermore, assume that both* $\|\boldsymbol{\theta}^*\|_0 = o(\sqrt{n}/\log(p)/\log(n))$ *and* $\|\boldsymbol{\gamma}^*\|_0 = o(\sqrt{n}/\log(p)/\log(n))$ *and that the normalization* $\|\boldsymbol{\gamma}^*\|_2 = \mathcal{O}(1)$ *and* $\|\boldsymbol{\theta}^*\|_2 = \mathcal{O}(1)$ *holds. Then, under the alternative hypothesis* $H_1$ *in (14), it holds that with* $n \to \infty$

$$T_n \stackrel{d}{=} \mathcal{N}(0, 1) + h\sigma_u\sigma_\epsilon^{-1}\frac{n^{-1}\mathrm{trace}(\widetilde{\boldsymbol{P}})}{\sqrt{n^{-1}\mathrm{trace}(\widetilde{\boldsymbol{P}}\boldsymbol{P}^{-1}\widetilde{\boldsymbol{P}})}} + o_P(1).$$

Theorem 4 establishes the power properties of our test and requires both $\boldsymbol{\gamma}^*$ and $\boldsymbol{\theta}^*$ to be sparse with a $o(\sqrt{n}/\log(p)/\log(n))$ sparsity rate. The deviation term depends on the choice of the matrix $\widetilde{\boldsymbol{P}}$ and by the Cauchy-Schwarz inequality it can be shown that

$$\frac{n^{-1}\mathrm{trace}(\widetilde{\boldsymbol{P}})}{\sqrt{n^{-1}\mathrm{trace}(\widetilde{\boldsymbol{P}}\boldsymbol{P}^{-1}\widetilde{\boldsymbol{P}})}} \leq \sqrt{n^{-1}\mathrm{trace}(\boldsymbol{P})}.$$

This implies that using the proxy matrix $\widetilde{\boldsymbol{P}}$ instead of the true unknown matrix $\boldsymbol{P}$ leads to an asymptotic loss of power.

If we consider the particular, naive choice of the proxy matrix $\boldsymbol{M} = \boldsymbol{0}_{Nq\times Nq}$ then $\widetilde{\boldsymbol{P}}$ reduces to the identity matrix $\boldsymbol{I}_{Nq}$. In that case the method reduces to the method entirely based on simple linear model and leads to a significant loss of power. For a simple choice of the matrix $\boldsymbol{M} = c\boldsymbol{I}_{Nq}$ where $c$ is a constant positive number, the matrix $\widetilde{\boldsymbol{P}}$ takes the form $(\boldsymbol{I}_n + c\boldsymbol{W}\boldsymbol{W}^\top)^{-1}$.

Using Theorem 4 we can explicitly derive the expression for power of the $T_n$ statistics under nominal level $\alpha$ as

$$\text{Power} = 2 - \Phi\left(\Phi^{-1}(1-\frac{\alpha}{2}) - D_n(h)\right) + \Phi\left(\Phi^{-1}(1-\frac{\alpha}{2}) + D_n(h)\right)$$

where $D_n(h) = h\sigma_u\sigma_\epsilon^{-1}\frac{n^{-1}\text{trace}(\widetilde{\boldsymbol{P}})}{\sqrt{n^{-1}\text{trace}(\widetilde{\boldsymbol{P}}\boldsymbol{P}^{-1}\widetilde{\boldsymbol{P}})}}$.

Furthermore, the length of the $1-\alpha$ confidence interval of $\beta^*$ constructed from Theorem 3 is given by $2n^{-1/2}|h_\alpha|$ for $h_\alpha$ satisfying $2\Phi\left(\Phi^{-1}(1-\alpha/2) - D_n(h_\alpha)\right) = 1-\alpha$ which can easily be estimated numerically.

We also note that the result can be extend to include fully dense models where $\|\boldsymbol{\gamma}^*\|_\infty$ is not allowed to grow rapidly with $n$ but the $l_0$ norm can be as large as $p$; or mixture settings with many small elements and only a small number of strong elements (here the sum of the small elements would be allowed to grow with $p$). We conjecture that the power function will look the same as it does in Theorem 4.

# 4 Numerical Experiments

In observational studies, accurate construction of p-values requires to overcome three potential sources of bias. First, there is the initial linear mixed model and in particular the size of the sparsity of the nuisance parameters $\boldsymbol{\gamma}^*$. We need to make sure that our test is reasonably stable whatever the structure of $\boldsymbol{\gamma}^*$ is. Then, second, there is the underlying precision matrix of the design matrix. Here a bias can be introduced with presence of large feature correlations and/or dependencies. Third, we need to make sure that the test is reasonably stable regardless of the exact structure of the random effects; we need to make sure that the test is not biased by changing matrix $\boldsymbol{\Psi}$, i.e. $\boldsymbol{\psi}$. The simulations here aim to test the ability of the introduced test to respond to all three of these factors.

## 4.1 Models and designs

We consider the mixed model

$$\boldsymbol{Y} = \boldsymbol{X}\boldsymbol{\beta}^* + \boldsymbol{W}\boldsymbol{b} + \epsilon$$

with $n = 200$ observations coming from $N = 50$ different groups of size $n_i = 4$. The vector $\boldsymbol{\beta}^*$ of fixed effects is of length $p = 500$. The components of $\boldsymbol{\epsilon}$ are independent and come from a standard Gaussian distribution. The random effect vector is defined as $\boldsymbol{b} = (\boldsymbol{b}_1^\top, \dots, \boldsymbol{b}_N^\top)^\top \in \mathbb{R}^{qN}$ with $\boldsymbol{b}_i \sim \mathcal{N}_q(0, \boldsymbol{\psi})$ for $i = 1, \dots, N$. We define $\boldsymbol{W} = \text{diag}(\boldsymbol{W}_1, \dots, \boldsymbol{W}_N) \in \mathbb{R}^{n\times qN}$ with the stack matrix $(\boldsymbol{W}_1^\top, \dots, \boldsymbol{W}_N^\top)^\top \in \mathbb{R}^{n\times q}$ consisting of the first $q$ columns of $\boldsymbol{X}$. We consider three models:

- *Model 1*: the rows of $\boldsymbol{X}$ come from a multivariate normal distribution with zero mean and Toeplitz covariance matrix $\boldsymbol{\Sigma}$ with $\boldsymbol{\Sigma}_{ij} = (-0.5)^{|i-j|}$. The components of $\boldsymbol{\beta}^*$ are generated as follows: $\boldsymbol{\beta}^* = 5\boldsymbol{a}/\left\|\boldsymbol{a}\right\|_2$ with $\boldsymbol{a} = (a_1, \dots, a_p)$, where $a_j$ is generated from a uniform distribution on $(0, 1)$ if $j \le 3s/2$ and $j/3$ is not an integer; otherwise, $a_j = 0$. We then have $\|\boldsymbol{\beta}\|_0 = s$. The model contains $q = 2$ random effects with covariance matrix $\boldsymbol{\psi} = \text{diag}(0.56, 0.56)$.

- *Model 2*: the setting is the same as in Model 1 except that $q = 3$ and $\boldsymbol{\psi} = \mathrm{diag}(3, 3, 2)$.
- *Model 3*: the setting is the same as in Model 1 except for $\boldsymbol{\Sigma}$ which is such that $\boldsymbol{\Sigma}_{ij} = 1$ if $i = j$, $\boldsymbol{\Sigma}_{ij} = -\rho/(1 + \rho^2)$ if $|i - j| = 1$, and $\boldsymbol{\Sigma}_{ij} = 0$ otherwise, where $\rho$ varies.

As noted in the appendix, the Toeplitz design implies that the underlying correlation model (4) is sparse with two non-zero components being equal (for $\rho = -0.5$) to $-0.4$. Thus, for models 1 and 2, the vector $\boldsymbol{\theta}^*$ is sparse. Conversely, model 3 is such that that the rows of $\boldsymbol{\Sigma}$ are sparse, while the row of its inverse are not sparse sparse. This in turn implies that $\boldsymbol{\theta}^*$ in (4) is a dense vector. Regarding the random effect part, similar choices of $q$ and $\boldsymbol{\psi}$ are made by Schelldorfer et al. (2011).

## 4.2 Tuning parameters

Tuning parameters need to be chosen in optimization problems (10) and (12). Based on our investigation, we suggest the following choices. The sensitivity of our procedure to the choice of the tuning parameters is given in Table 4. Similarly to Zhu and Bradic (2016), we choose $\bar{\eta}_\gamma = 0.05\, \boldsymbol{V}^\top \widetilde{\boldsymbol{P}} \boldsymbol{V}/n$. We define $\eta_\gamma = \sqrt{0.5 n^{-1} \log p}\ \widehat{\sigma}$ and $\mu_\gamma = 4\sqrt{\log n}\ \widehat{\sigma}$ with $\widehat{\sigma} = \left\| \widetilde{\boldsymbol{P}}(\boldsymbol{V} - \boldsymbol{X}\boldsymbol{\gamma}_{\mathrm{init}}) \right\|_2$ where $\boldsymbol{\gamma}_{\mathrm{init}}$ is obtained by a linear estimation (without the random effects) with the scaled lasso. Such an initial estimator of $\boldsymbol{\gamma}$ is also used by Rohart et al. (2014). The choice of the tuning parameters $\bar{\eta}_\theta$, $\eta_\theta$, $\mu_\eta$ and $\eta'_\theta$ is done in a similar way.

Regarding the choice of the matrix $\widetilde{\boldsymbol{P}} = (\boldsymbol{I}_n + \boldsymbol{W}\boldsymbol{M}\boldsymbol{W}^\top)^{-1}$ serving as a proxy of $\boldsymbol{P} = (\boldsymbol{I}_n + \sigma_\epsilon^{-2}\boldsymbol{W}\boldsymbol{\Psi}\boldsymbol{W}^\top)^{-1}$, we show in Lemma 4 that there is a large pool of possible choices guaranteeing good asymptotic results. Fan and Li (2012) suggest to use $\boldsymbol{M} = (\log n)\boldsymbol{I}_{Nq}$. In the present simulation study we use the slightly more elaborate $\boldsymbol{M} = \sigma_{\epsilon,\mathrm{init}}^{-2}\mathrm{diag}(\boldsymbol{\psi}_{\mathrm{init}}, \ldots, \boldsymbol{\psi}_{\mathrm{init}})$ where initial estimators are derived similarly as in Rohart et al. (2014): using $\boldsymbol{\gamma}_{\mathrm{init}}$ obtained by a linear model estimation with the scaled lasso, we compute $\sigma_\epsilon^{2[-1]} = \frac{1}{n-\mathrm{df}} \|\boldsymbol{V} - \boldsymbol{X}\boldsymbol{\gamma}_{\mathrm{init}}\|_2^2$, $\psi_{\mathrm{init}} = \frac{0.4}{q}\sigma_\epsilon^{2[-1]}\boldsymbol{I}_q$ and $\sigma_{\epsilon,\mathrm{init}}^2 = 0.6\sigma_\epsilon^{2[-1]}$.

## 4.3 Numerical results

We test $H_{0,j} : \beta_j^* = \beta_{0,j}$ for $j = 3, 4, 100$ and study the behaviour of our procedure. These values of $j$ correspond to different types of coefficients; components 3 and 100 are not active while component 4 is active. Furthermore, in the Toeplitz case, component 100 is almost not correlated with any of the active variables. We set $H_{1,j} : \beta_j^* = \beta_{0,j} + hn^{-1/2}$ where the situation $h = 0$ corresponds to size properties while $h \neq 0$ corresponds to power properties. All our results are based on 1000 replications and on a 5% nominal level. To the best of our knowledge, our procedure is the first one to do hypothesis testing for high-dimensional mixed models which makes it hard to find competitors. We compare our mixed model method to the linear model method of Zhu and Bradic (2016).

### 4.3.1 Gaussian mixed model

In Figure 1, we report size and power properties of our procedure for Model 1 and for different components of $\boldsymbol{\beta}^*$. On Figure 1 (a) we observe that the false discovery rate is close to the 5% nominal level whatever the sparsity level. Our procedure can handle dense $\boldsymbol{\beta}^*$ and this is one of its major features. Sparsity is required only for either $\boldsymbol{\theta}^*$ or $\boldsymbol{\beta}^*$ to obtain reliable results under the null hypothesis. Note that under the alternative hypothesis $h \neq 0$, our theoretical results require both of them to be sparse, see Theorem 4. On Figure 1 (b), we provide the power of our test for Model 1 in a sparse model ($s = 5$). Interestingly, the probability to reject the null hypothesis is not symmetric in $h$. This can be explained by the fact that we use a penalized estimator which shrinks coefficients towards zero. The correlation between the components compensates (here for $h$ negative) or amplifies (here for $h$ positive) the bias of the estimator, which leads to an asymmetry of the results.

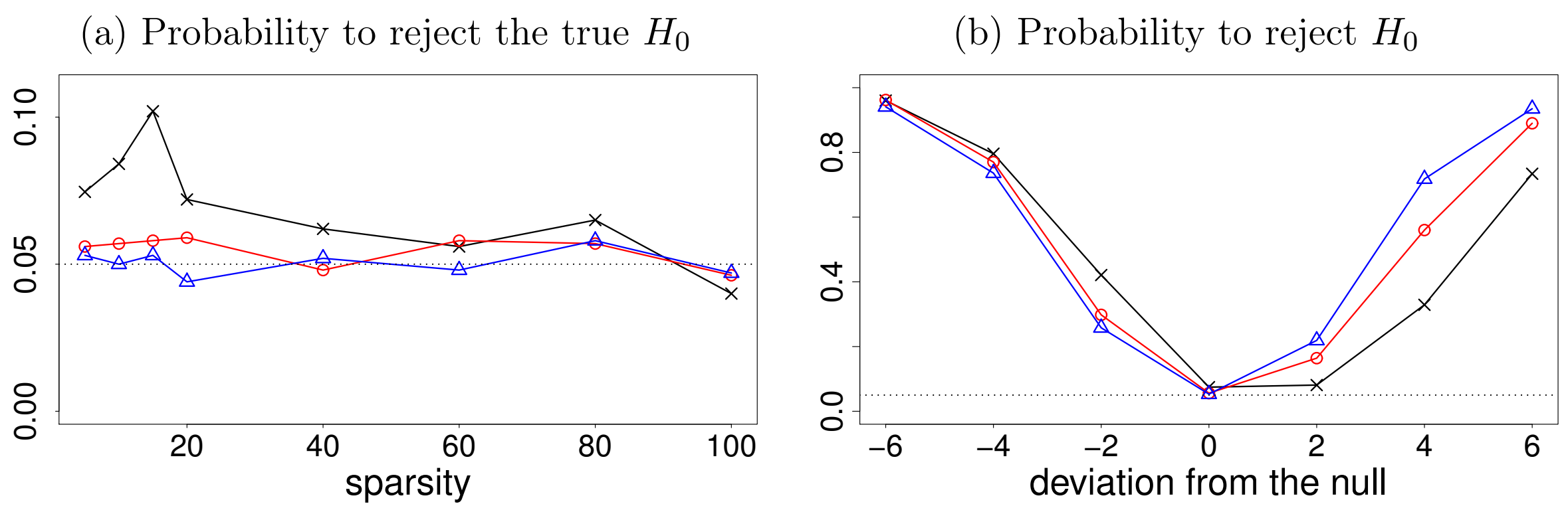

Figure 1: Empirical rate of rejection of the null hypothesis under (a) the null hypothesis, and (b) alternative hypothesis, at a 5% nominal level for Model 1. As described in the text, hypothesis tests are performed on different components of $\beta^*$: the third component (black crosses), the fourth component (red circles), and the 100-th component (blue triangles).

In Figure 2 we give size and power properties of our procedure for Model 3 in which $\boldsymbol{\theta}^*$ is dense, i.e. $s_\theta = p$ in this particular case. We vary the size of the correlation $\rho$ with larger values corresponding to larger size of the coefficients of $\boldsymbol{\theta}^*$. We observe that the proposed test is remarkably resilient and controls Type I error in finite samples. We observe that when $\boldsymbol{\gamma}^*$ is not sparse, which here corresponds to sparsity larger than 5 (note that $\sqrt{n}/\log(p) \approx 5$ for $n = 200$, $p = 500$), our method is not guaranteed to control Type I error; however, we see that our method strikingly controls errors even in such cases whenever $\rho < 0.5$. It is worth pointing that no method is expected to perform well when all elements are dense and correlation is high. Hence, we take this behavior to be close to optimal.

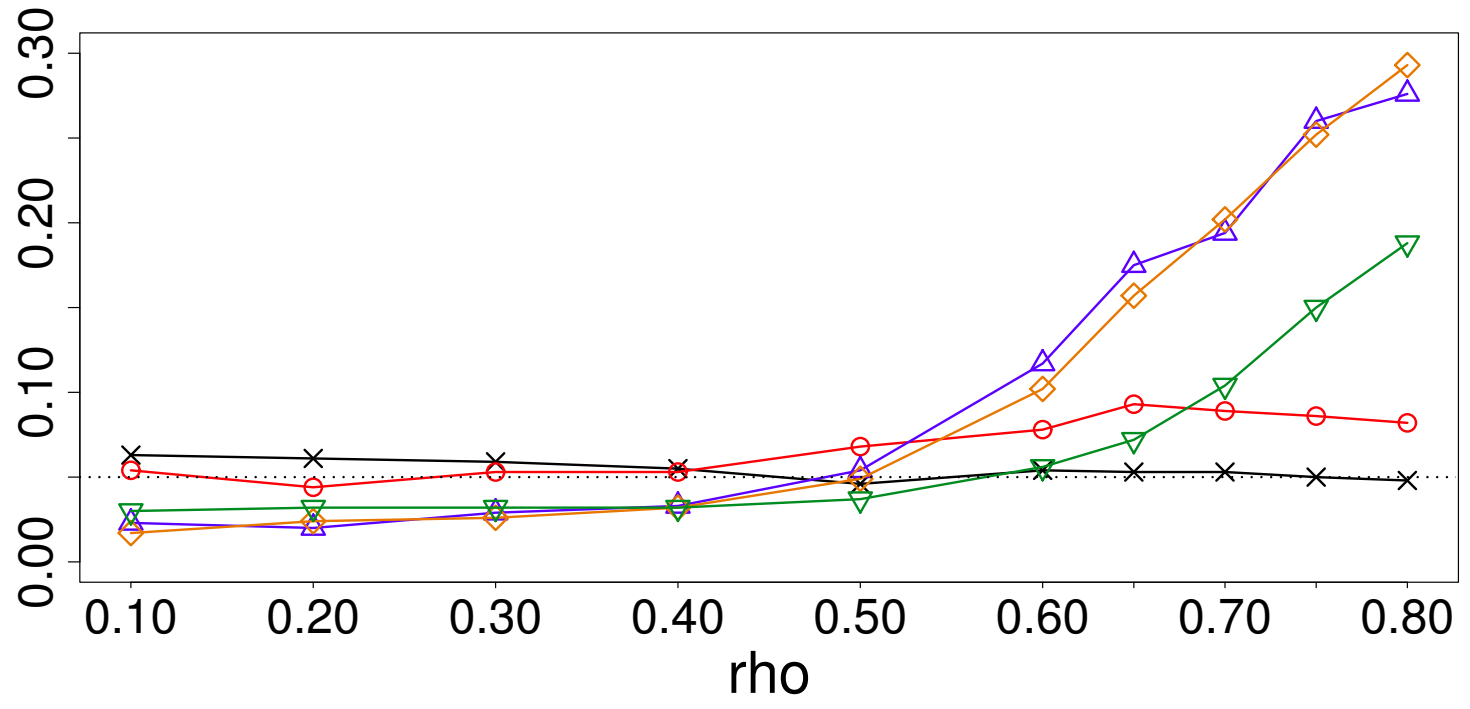

Figure 2: Empirical rate of rejection of the true null hypothesis at a 5% nominal level for Model 3 with dense vector $\boldsymbol{\theta}^*$. We vary the level of sparsity of $\boldsymbol{\gamma}^*$ and correlation level $\rho$. Hypothesis testing is performed on the second component of $\boldsymbol{\beta}^*$. Empirical Type I errors are plotted for different sparsity levels: black crosses 2, red circles 3, blue triangles 4, orange diamonds 5, green upside-down triangles 10.

In Table 2, we compare our method, for model 1 and 2, to the method of Zhu and Bradic (2016) designed for linear models and which thus does not take into account the group structure of the mixed model. For both models, the probability to reject the null hypothesis is closer to the 5% nominal level when the null hypothesis holds ($h = 0$) and is larger under the alternative hypothesis ($h \neq 0$), compared to the linear model procedure.

| | Probability to reject the null hypothesis at a 5% level | | | |
|---|---|---|---|---|
| | Model 1 | | Model 2 | |
| | LM procedure | MM procedure | LM procedure | MM procedure |
| $h = -6$ | 0.85 | 0.96 | 0.46 | 0.65 |
| $h = -4$ | 0.60 | 0.77 | 0.30 | 0.39 |
| $h = -2$ | 0.29 | 0.30 | 0.19 | 0.16 |
| $h = 0$ | 0.11 | 0.06 | 0.12 | 0.05 |
| $h = 2$ | 0.11 | 0.16 | 0.07 | 0.05 |
| $h = 4$ | 0.28 | 0.56 | 0.07 | 0.18 |
| $h = 6$ | 0.55 | 0.89 | 0.13 | 0.42 |

Table 2: Probability of rejecting the null hypothesis as a function of the deviation from the null. Nominal level is taken to be 5%. *LM procedure* consists in applying the linear model procedure of Zhu and Bradic (2016) to the mixed model while *MM procedure* consists in applying our mixed model procedure. Parameters are $n = 200$, $p = 500$, $N = 50$, $s = 5$. The 4th component of $\boldsymbol{\beta}^*$ is tested. Model 1 contain 2 random effects and Model 2 contains 3 random effects with large magnitude.

#### 4.3.2 Model with heavy-tailed design

In the next example we consider a model that departs from normality assumptions. Parameters choices are done in the same way as in Model 1: $n = 200$, $N = 50$, $p = 500$, and $q = 2$.

- *Model 4*: the setting is the same as in Model 1 except that the entries of $\boldsymbol{\Sigma}^{-1/2}x_i$ and of $\boldsymbol{\epsilon}$ are generated from a student $t$ distribution with 10 degrees of freedom.
- *Model 5*: the setting is the same as in Model 1 except that the entries of $\boldsymbol{\Sigma}^{-1/2}x_i$ and of $\boldsymbol{\epsilon}$ are generated from a student $t$ distribution with 3 degrees of freedom.

We perform hypothesis test for the fourth component of $\boldsymbol{\beta}^*$ for Models 1, 4, and 5 and report empirical sizes and powers in Figure 3. We observe empirical coverages close to the 5% nominal level for the three models under the null hypothesis whatever the sparsity level is. We also observe that the presence of heavier tailed errors results in a decrease of power.

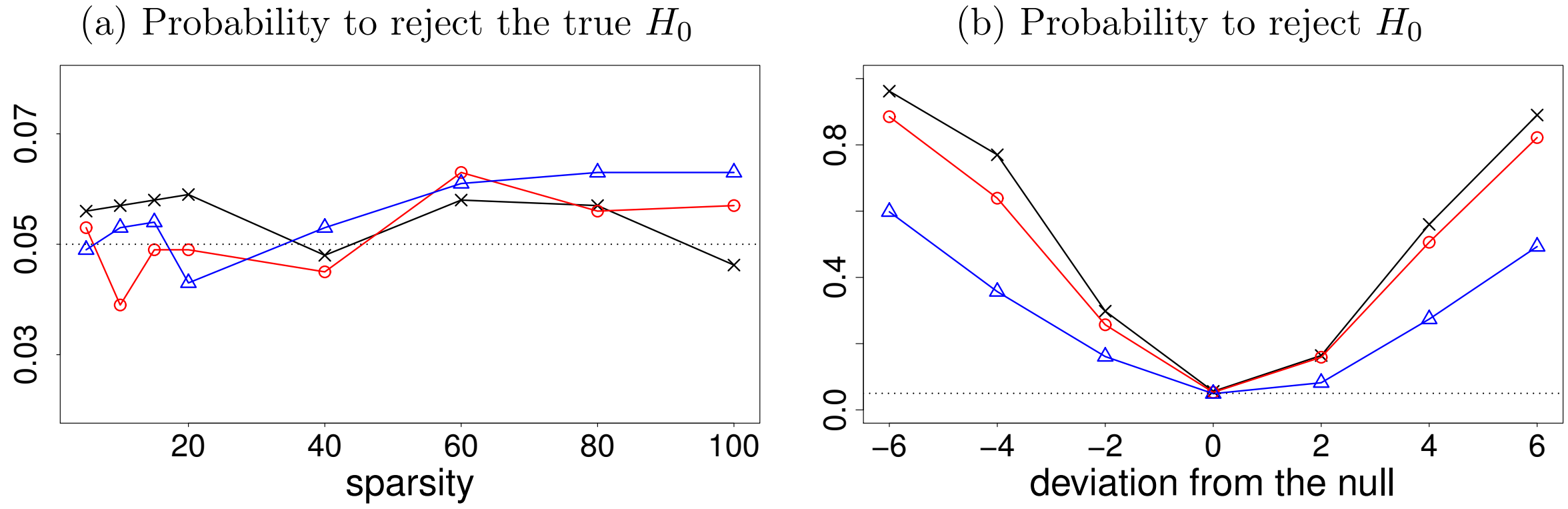

Figure 3: Empirical rate of rejection of the null hypothesis under (a) the null hypothesis, and (b) alternative hypothesis, at a 5% nominal level for different distributions of the error: Gaussian (model 1, black crosses), Student with 10 degrees of freedom (model 4, red circles), and Student with 3 degrees of freedom (model 5, blue triangles). The hypothesis test is the $H_{0,4} : \beta_4^* = \beta_{0,4}$ versus $H_{1,4} : \beta_4^* = \beta_{0,4} + hn^{-1/2}$.

#### 4.3.3 Choice of the proxy matrix

The choice of the proxy matrix $\widetilde{\boldsymbol{P}} = (\boldsymbol{I}_n + \boldsymbol{W}\boldsymbol{M}\boldsymbol{W}^\top)^{-1}$ has an influence on the power of the test. Our default choice, presented in section 4.2 and inspired by Rohart et al. (2014) consists in working with $\boldsymbol{M} = \sigma_{\epsilon,\text{init}}^{-2}\text{diag}(\boldsymbol{\psi}_{\text{init}}, \ldots, \boldsymbol{\psi}_{\text{init}})$. With these initial estimators, this is equivalent to work with $M = \frac{2}{3q}\boldsymbol{I}_{Nq}$. A second choice, proposed by Fan and Li (2012), is to work with $\boldsymbol{M} = \log n \boldsymbol{I}_{Nq}$. Simulation results for those choices of the proxy matrix are reported in Table 3 for models 1 and 2. We first observe that the two choices provide the 5% nominal level under the null hypothesis. Secondly, we observe slight differences in terms of power which can be explained by how well the true variance structure is approximated by the proxy matrix $M$.

| | Probability to reject the null hypothesis at a 5% level | | | |
|---|---|---|---|---|
| | Model 1 | | Model 2 | |
| | $\boldsymbol{M} = \sigma_{\epsilon,\text{init}}^{-2}\boldsymbol{\psi}_{\text{init}}$ | $\boldsymbol{M} = \log n \boldsymbol{I}_{Nq}$ | $\boldsymbol{M} = \sigma_{\epsilon,\text{init}}^{-2}\boldsymbol{\psi}_{\text{init}}$ | $\boldsymbol{M} = \log n \boldsymbol{I}_{Nq}$ |
| $h = -6$ | 0.96 | 0.91 | 0.65 | 0.79 |
| $h = -4$ | 0.77 | 0.66 | 0.39 | 0.49 |
| $h = -2$ | 0.30 | 0.25 | 0.16 | 0.19 |
| $h = 0$ | 0.06 | 0.05 | 0.05 | 0.05 |
| $h = 2$ | 0.16 | 0.15 | 0.05 | 0.09 |
| $h = 4$ | 0.56 | 0.49 | 0.18 | 0.30 |
| $h = 6$ | 0.89 | 0.85 | 0.42 | 0.64 |

Table 3: Probability to reject the null hypothesis in function of the deviation from the null for a 5% nominal level, and in function of the choice of the proxy matrix. Parameters are $n = 200$, $p = 500$, $N = 50$, $s = 5$. The 4th component of $\boldsymbol{\beta}^*$ is tested. Model 1 contains 2 random effects and Model 2 contains 3 random effects with large magnitude.

#### 4.3.4 Sensitivity to tuning parameters choice

We conclude this simulation study by an analysis of the sensitivity of our procedure to the choice of the tuning parameters. We consider Model 1 and two scenarios: $(h, s) = (0, 40)$ and $(h, s) = (4, 5)$ corresponding to a non-sparse model under $H_0$ and an sparse model under the true alternative $H_1$. We perform hypothesis tests for the fourth component of $\boldsymbol{\beta}$ and provide results for different choices of the tuning parameters. Results of Table 4 illustrate that the method is reasonably weakly sensitive to the tuning parameters choice.

| | Probability to reject $H_0$ at a 5% level | | | | | |
|---|---|---|---|---|---|---|
| | $h = 0$ and $s = 40$ | | | $h = 4$ and $s = 5$ | | |
| | $\frac{1}{2}\eta$ | $\eta$ | $2\eta$ | $\frac{1}{2}\eta$ | $\eta$ | $2\eta$ |
| $0.5\mu$ | 0.06 | 0.05 | 0.05 | 0.48 | 0.48 | 0.53 |
| $\mu$ | 0.04 | 0.05 | 0.05 | 0.59 | 0.59 | 0.59 |
| $2\mu$ | 0.06 | 0.05 | 0.06 | 0.35 | 0.34 | 0.34 |

| | Probability to reject $H_0$ at a 5% level | |
|---|---|---|
| | $h = 0$ and $s = 40$ | $h = 4$ and $s = 5$ |
| $\frac{1}{2}\bar{\eta}$ | 0.04 | 0.60 |
| $\bar{\eta}$ | 0.05 | 0.59 |
| $2\bar{\eta}$ | 0.04 | 0.42 |

Table 4: Probability to reject the null hypothesis for different choices of the tuning parameters and for Model 1 with $j = 4$ and $(h, s) = (0, 40)$ or $(h, s) = (4, 5)$, compared with the default choice presented in Section 4.2. Notation $\eta$ refers to $\eta_\gamma$, $\eta_\theta$ and $\eta'_\theta$; $\mu$ refers to $\mu_\gamma$ and $\mu_\theta$; $\bar{\eta}$ refers to $\bar{\eta}_\gamma$ and $\bar{\eta}_\theta$. If it is not specified, the default choice is used.

### 4.4 Hypothesis testing for Riboflavin data

We study the Riboflavin data which contains 4088 gene expressions from 111 observations divided in 28 groups from size 2 to 6, and which has also been considered by Schelldorfer et al. (2011). The response variable is the logarithm of the riboflavin production rate of Bacillus subtilis. In the context of linear models a related dataset was considered in van de Geer et al. (2014) and Javanmard and Montanari (2014b). The data of our interest has a clear group structure, and therefore we include a random intercept in the linear model, resulting into the following mixed model

$$\boldsymbol{Y} = \boldsymbol{X}\boldsymbol{\beta}^* + \boldsymbol{W}\boldsymbol{b} + \boldsymbol{\epsilon}$$

where $W$ contains 1 in the appropriates entries, $\boldsymbol{Y}$ is of length 111 and $\boldsymbol{X}$ is the $111 \times 4089$ design matrix with its first column containing only ones and its other columns corresponding to the 4088 standardized covariates. The vector $\boldsymbol{b}$ contains the 28 realizations of the random effects. Its components are independent and identically distributed with mean zero and unknown variance $\boldsymbol{\psi}$.

Riboflavin (vitamin B2) is an essential component of the basic metabolism. The riboflavin biosynthesis in bacteria was analyzed from a biological perspective using comparative analysis of genes, operons and regulatory elements. However, little is known about the mechanisms of regulation of the bacterial riboflavin genes.

Gene YpaA has been verified experimentally to be a transporter of riboflavin or related compounds, co-regulated with other riboflavin genes (Vitreschak et al., 2002). The RFN element (specific locus related to RNA folding) was only encoded upstream of the YpaA gene. Moreover, in B. subtilis, riboflavin uptake was increased when YpaA was over-expressed and abolished when YpaA was deleted (Vogl et al., 2007). Hence it makes sense to test a one sided hypothesis with the null being that the corresponding $\beta^*$ is smaller or equal to zero and alternative that it is larger than zero. In this context, we have applied our test above while including all of the 4087 remaining genes in the study. We have obtained a test statistic value of 1.60. Therefore our method is able to identify YpaA gene at a 6% significance level; we should note that our sample size is extremely small and deviations from 5% should be expected due to an extremely small sample size. It is also worth noting, that no previous study of this dataset was able to identify YpaA as statistically significant gene. We also have observed large correlation among the genes present in this study, which may additionally explain why previous methods failed; observe that our method is doubly robust and is able to overcome high correlations in the data.

The best studied system of the riboflavin biosynthesis in bacteria is the *rib* operon of Bacillus subtilis. Our data contains information about the Bacillus subtilis and there are a number of genes (but not all) related to *rib* operon. We are particularly interested in the ribB gene which was reported as over-expressed in a number of chemical and biological studies (Mörtl et al., 1996) but has not been yet studied from a statistical point of view. We have performed a single hypothesis test related to the $\beta^*$ corresponding to ribB gene; while including all of the remaining genes. We have found that our test statistic is able to detect ribB with a 1% nominal value; the observed value of the test statistics is 2.69. This result confirms the biological evidence and reconfirms wide-applicability of the proposed test.

Another part of the *rib* operon is a ribC gene. The ribC gene was cloned and sequenced and it was determined that it plays an essential role in the flavin metabolism of B. subtilis (Mack et al., 1998). In particular it was observed that it is over-expressed and that it suppresses the riboflavin overproduction. We have performed an one-sided hypothesis test to observe statistical significance of this gene in the current dataset. We have observed a test statistic value of 1.01 leading to a 16% significance level. Although larger than 10%, this study is one of the first statistical studies that was able to detect ribC gene at any level. For completeness, we also report that an averaged size of the selected sets in the original and the feature model was 17.90 and 741.10.

## 5 Further explorations

The methodology presented in the previous sections is extremely general and provides easy extensions to a number of interesting problems and settings.

### 5.1 Power improvements

Our test statistic uses a proxy matrix $\widetilde{\boldsymbol{P}}$ which has some influence on the power through the deviation term $h\sigma_u\sigma_\epsilon^{-1}\frac{n^{-1}\mathrm{trace}(\widetilde{\boldsymbol{P}})}{\sqrt{n^{-1}\mathrm{trace}(\widetilde{\boldsymbol{P}}\boldsymbol{P}^{-1}\widetilde{\boldsymbol{P}})}}$ in Theorem 4. This deviation term is smaller than $\sqrt{n^{-1}\mathrm{trace}(\boldsymbol{P})}$ and it is expected that the "closer" $\widetilde{\boldsymbol{P}}$ is to $\boldsymbol{P}$, the bigger this deviation term is and thus the better the power of the test statistics is. By using a good estimator $\widehat{\boldsymbol{P}}$ of $\boldsymbol{P}$, we could expect the deviation term to be close to $h\sigma_u\sigma_\epsilon^{-1}\sqrt{n^{-1}\mathrm{trace}(\boldsymbol{P})}$.

This leads to the following scheme:

Step 1: Compute an estimator $\widehat{\boldsymbol{\gamma}}$ by solving (10) with the proxy matrix $\widetilde{\boldsymbol{P}}$. We know by Lemma 6 that $\|\boldsymbol{\gamma}-\boldsymbol{\gamma}^*\|_1$ is small.

Step 2: Obtain a better estimator $\widehat{\boldsymbol{P}}$ of $\boldsymbol{P}$ by solving (17) .

Step 3: Compute a new estimator of $\boldsymbol{\gamma}^*$ obtained by solving (10) with $\widehat{\boldsymbol{P}}$ instead of $\widetilde{\boldsymbol{P}}$ and construct a $T_n$ statistics based on this new estimator and on $\widehat{\boldsymbol{P}}$.

As maximum likelihood estimators are often biased for the estimation of the variance components, we utilized the marginal log likelihood and propose to consider the following restricted maximum likelihood estimator

$$\widehat{\boldsymbol{P}} = \arg\min_{\boldsymbol{P}\geq 0}\left\{\frac{1}{2}(\boldsymbol{V}-\boldsymbol{X}\widehat{\boldsymbol{\gamma}})^\top\boldsymbol{P}(\boldsymbol{V}-\boldsymbol{X}\widehat{\boldsymbol{\gamma}})+\frac{1}{2}\log(\det(\boldsymbol{P}^{-1}))+\frac{1}{2}\log(\det(\boldsymbol{X}^\top\boldsymbol{P}\boldsymbol{X}))\right\}. \tag{17}$$

Theorems 3 and 4 are expected to hold with this new $T_n$ statistic.

With this approach, not only do we obtain a better test statistic (more efficient) but we are able to estimate the error variance of the initial model by a new estimator which can be defined as $\widehat{\sigma}_\epsilon^2 = n\widehat{\sigma}^2/\mathrm{trace}(\widehat{\boldsymbol{P}})$.

### 5.2 Multivariate Testing

For testing general multivariate hypotheses of the kind $H_0 : \boldsymbol{\beta}^* = \boldsymbol{\beta}_0$ versus $H_1 : \boldsymbol{\beta}^* \neq \boldsymbol{\beta}_0$ where $\boldsymbol{\beta}^* \in \mathbb{R}^d$ and $d \to \infty$ ($p \to \infty$) we can easily adapt the procedure of Section 2.

With a little abuse of notation let $\boldsymbol{V} = \boldsymbol{Y} - \boldsymbol{Z}\boldsymbol{\beta}_0$ denote the pseudo-response vector (similar to previous sections) where now a univariate parameter $\beta_0$ is replaced with a $d$-dimensional counterpart $\boldsymbol{\beta}_0$. Then with $\widehat{\boldsymbol{\gamma}}$ as defined before, we consider $d$ estimators $\widehat{\boldsymbol{\theta}}_{(1)}, \widehat{\boldsymbol{\theta}}_{(2)}, \cdots, \widehat{\boldsymbol{\theta}}_{(d)}$ defined as follows

$$\begin{aligned}\widehat{\boldsymbol{\theta}}_{(j)} \quad &\in \quad \arg\min_{\boldsymbol{\theta}_{(j)}\in\mathbb{R}^{p-d}} \|\boldsymbol{\theta}_{(j)}\|_1 \\ \text{such that} \quad & \|n^{-1}\boldsymbol{X}^\top(\boldsymbol{Z}_{(j)}-\boldsymbol{X}\boldsymbol{\theta}_{(j)})\|_\infty \leq \eta_{\theta,j} \\ & n^{-1}\boldsymbol{Z}_{(j)}^\top(\boldsymbol{Z}_{(j)}-\boldsymbol{X}\boldsymbol{\theta}_{(j)}) \geq \bar{\eta}_{\theta,j} \\ & \|\boldsymbol{Z}_{(j)}-\boldsymbol{X}\boldsymbol{\theta}_{(j)}\|_\infty \leq \mu_{\theta,j} \\ & \|n^{-1}\boldsymbol{X}^\top\widehat{\boldsymbol{P}}(\boldsymbol{Z}_{(j)}-\boldsymbol{X}\boldsymbol{\theta}_{(j)})\|_\infty \leq \eta_{\theta,j}\ .\end{aligned} \tag{18}$$

Then, we consider to reject the null hypothesis whenever $T_n = \max_j T_{n,j}$ is larger than the bootstrap quantile $q_{1-\alpha}$ to be defined below. The test statistics $T_{n,j}$ are defined in the same spirit of the previous section and take the form of

$$T_{n,j} = n^{-1/2} \sum_{i=1}^{n} T_{ij}, \qquad T_{ij} = \frac{(Z_{i(j)} - \boldsymbol{X}_i \widehat{\boldsymbol{\theta}}_{(j)})^\top \widetilde{\boldsymbol{P}} \left(v_i - \boldsymbol{X}_i \widehat{\gamma}\right)}{\widehat{\sigma}_{u,j} \widehat{\sigma}_\epsilon}$$

where $\widehat{\sigma}_\varepsilon^2 = n^{-1} \left\| \widetilde{\boldsymbol{P}} \left( \boldsymbol{V} - \boldsymbol{X} \widehat{\gamma} \right) \right\|_2^2$, $\widehat{\sigma}_{u,j}^2 = n^{-1} \| \boldsymbol{Z}_{(j)} - \boldsymbol{X} \boldsymbol{\theta}_{(j)} \|_2^2$.

The quantile $q_{1-\alpha}$ is defined as a $1-\alpha$ quantile of the distribution of a bootstrapped test statistics $\widetilde{T}_n = \max_j \widetilde{T}_{n,j}$ with

$$\widetilde{T}_{n,j} = n^{-1/2} \sum_{i=1}^{n} \xi_i (T_{ij} - n^{-1/2} T_{n,j})$$

for a class of multipliers $\{\xi_i\}_{i=1}^n$ that are drawn independently from the data and that follow a standard Gaussian distribution.

For a more general hypothesis $H_{0,j} : \beta_j^* = \beta_{0,j}$ for $j \in \mathcal{J}$, $\mathcal{J} \subset \{1, 2, \cdots, p\}$ we can consider the maximum as the test statistic and denote by $T_n(\beta_{0,j})$ the test statistics as in (13) where one element of $\beta^*$ is hold-out and the remaining ones are stacked in the vector $\gamma^*$. Then, we use $G_{\mathcal{J}}(c) = P\left[\max_{j \in \mathcal{J}} (|\widetilde{T}_{n,j}|) \le c\right]$, Then, the p-value for $H_{0,\mathcal{J}} : \beta^* = \beta_0$, against the alternative being the complement, is defined as $P_{\mathcal{J}} = 1 - G_{\mathcal{J}}(\max_{j \in \mathcal{J}} |T_n(\beta_{0,j})|)$).

## 5.3 Generalized Linear Mixed Models

We note that inference in generalized linear models is an extremely difficult problem, even in low-dimensional setting. Typical approaches are hindered by a difficult numerical integration and a often non-analytic expression of a likelihood or profile likelihood function. In this subsection we illustrate how the methodology introduced in Section 2 can be easily extended for this difficult setting. Robustness of our approach provides flexibility regarding specifying the likelihood exactly.

Let $\boldsymbol{Y}$ be the observed data vector and, conditional on the random effects, $\boldsymbol{b}$, assume that the elements of $\boldsymbol{Y}$ are independent and drawn from a distribution in the exponential family (which, for simplicity of exposition, we take with the canonical link). To complete the specification, assume a distribution for $\boldsymbol{b}$ to be dependent on variance parameters, $\boldsymbol{D}$:

$$f_{\mathbf{y}_i|\boldsymbol{b}}(\mathbf{y}|\boldsymbol{b}, \boldsymbol{\gamma}^*, \boldsymbol{\beta}^*, \phi) = \exp\left\{ \frac{\mathbf{y}\boldsymbol{\eta}_i - c(\boldsymbol{\eta}_i)}{a(\phi)} + d(\mathbf{y}, \phi) \right\} \tag{19}$$

where $\boldsymbol{\eta}_i = \boldsymbol{X}_i \boldsymbol{\gamma}^* + \boldsymbol{Z}_i \boldsymbol{\beta}^* + \mathbf{W}_i \boldsymbol{b}_i$. With $g$ denoting the link function, we have $b' = g^{-1}$

$$E[\mathbf{y}_i|\boldsymbol{b}] = b'(\boldsymbol{X}_i \boldsymbol{\gamma}^* + \boldsymbol{Z}_i \boldsymbol{\beta}^* + \mathbf{W}_i \boldsymbol{b}_i).$$

We propose the following test statistic for use in generalized linear mixed models

$$T_{n,j} = \frac{n^{-1/2} (\boldsymbol{Z}_{(j)} - \boldsymbol{X} \widehat{\boldsymbol{\theta}}_{(j)})^\top \widetilde{\boldsymbol{P}} \left( \boldsymbol{Y} - b'(\boldsymbol{X} \widehat{\boldsymbol{\gamma}} + \boldsymbol{Z} \boldsymbol{\beta}_0) \right)}{\widehat{\sigma}_{u,j} \widehat{\sigma}_\epsilon}$$

with $\widehat{\sigma}_\varepsilon^2 = n^{-1} \left\| \widetilde{\boldsymbol{P}} \left( \boldsymbol{Y} - b'(\boldsymbol{X} \widehat{\gamma} + \boldsymbol{Z} \boldsymbol{\beta}_0) \right) \right\|_2^2$, $\widehat{\sigma}_{u,j}^2 = n^{-1} b''(\boldsymbol{X} \widehat{\gamma} + \boldsymbol{Z} \boldsymbol{\beta}_0) \| \boldsymbol{Z}_{(j)} - \boldsymbol{X} \widehat{\boldsymbol{\theta}}_{(j)} \|_2^2$.

For the procedure to be adaptive to generalized linear models, the estimators of $\boldsymbol{\gamma}^*$ and $\boldsymbol{\theta}_{(j)}^*$ need to be carefully developed. Regarding the estimation of $\boldsymbol{\gamma}^*$ we adapt the estimator of Section 2.2 and propose the following estimator

$$\begin{aligned} \widehat{\boldsymbol{\gamma}} \in \quad & \arg\min_{\boldsymbol{\gamma}\in\mathbb{R}^{p-d}} \|\boldsymbol{\gamma}\|_1 \\ \text{such that} \quad & \left\| n^{-1}\boldsymbol{X}^\top \widetilde{\boldsymbol{P}}\Big(\boldsymbol{Y} - b'(\boldsymbol{X}\boldsymbol{\gamma} + \boldsymbol{Z}\boldsymbol{\beta}_0)\Big)\right\|_\infty \le \eta_\gamma \\ & n^{-1}\boldsymbol{Y}^\top \widetilde{\boldsymbol{P}}\Big(\boldsymbol{Y} - b'(\boldsymbol{X}\boldsymbol{\gamma} + \boldsymbol{Z}\boldsymbol{\beta}_0)\Big) \ge \bar{\eta}_\gamma \\ & \|\widetilde{\boldsymbol{P}}(\boldsymbol{Y} - b'(\boldsymbol{X}\boldsymbol{\gamma} + \boldsymbol{Z}\boldsymbol{\beta}_0))\|_\infty \le \mu_\gamma \end{aligned} \tag{20}$$

for suitable choices of tuning parameters $\eta_\gamma \asymp \sqrt{n^{-1}\log(p)}$, $0 < \bar{\eta}_\gamma < \sigma_\varepsilon^2$ and $\mu_\gamma \asymp \sqrt{\log(n)}$. The estimator for $\boldsymbol{\theta}^*_{(j)}$ can now be defined with

$$\begin{aligned} \widehat{\boldsymbol{\theta}}_{(j)} \quad \in \quad & \arg\min_{\boldsymbol{\theta}_{(j)}\in\mathbb{R}^{p-d}} \|\boldsymbol{\theta}_{(j)}\|_1 \\ \text{such that} \quad & \|n^{-1}b''(\boldsymbol{X}\widehat{\boldsymbol{\gamma}} + \boldsymbol{Z}\boldsymbol{\beta}_0)\boldsymbol{X}^\top(\boldsymbol{Z}_{(j)} - \boldsymbol{X}\boldsymbol{\theta}_{(j)})\|_\infty \le \eta_{\theta,j} \\ & \|n^{-1}b''(\boldsymbol{X}\widehat{\boldsymbol{\gamma}} + \boldsymbol{Z}\boldsymbol{\beta}_0)\boldsymbol{X}^\top \widetilde{\boldsymbol{P}}(\boldsymbol{Z}_{(j)} - \boldsymbol{X}\boldsymbol{\theta}_{(j)})\|_\infty \le \bar{\eta}_{\theta,j} \\ & n^{-1}b''(\boldsymbol{X}\widehat{\boldsymbol{\gamma}} + \boldsymbol{Z}\boldsymbol{\beta}_0)\boldsymbol{Z}_{(j)}^\top(\boldsymbol{Z}_{(j)} - \boldsymbol{X}\boldsymbol{\theta}_{(j)}) \ge \bar{\eta}_{\theta,j} \\ & \|\boldsymbol{Z}_{(j)} - \boldsymbol{X}\boldsymbol{\theta}_{(j)}\|_\infty \le \mu_{\theta,j}, \end{aligned} \tag{21}$$

for suitable choices of tuning parameters $\eta_{\theta,j} \asymp \sqrt{n^{-1}\log(p)}$, $0 < \bar{\eta}_{\theta,j} < \sigma_{u,j}^2$, and $\mu_{\theta,j} \asymp \sqrt{\log(n)}$. Here, $\boldsymbol{Z}_{(j)} = (z_{1,(j)}, \cdots, z_{n,(j)})^\top \in \mathbb{R}^n$. Observe that differently from the linear mixed models, in the case of the generalized linear mixed models, the two estimators $\widehat{\boldsymbol{\gamma}}$ and $\widehat{\boldsymbol{\theta}}_{(j)}$ are dependent. The above procedure can be solved using iterations of linear programs, much in the spirit of weighted least squares methods. The final test statistic is now defined in a similar manner as before with $T_n = \max_j T_{n,j}$ where the multiplier bootstrap of the previous subsection can be successfully applied.

# 6 Discussion

This paper proposed a class of test statistics for performing inference on fixed effects that allow for high-dimensional and misspecified linear mixed models all while maintaining the benefits of classical methods, i.e., an asymptotically normal and unbiased test statistic with valid and honest confidence intervals. Our test statistics can be thought of as a doubly robust approach; it adapts to the sparsity of the nuisance parameters (fixed) as well as the unknown structure of the random effects (both variance and distributions). Such adaptivity seems essential for modern large-scale applications with many features, various sparsity assumptions as well as distributional assumptions that cannot be checked.

In general, the challenge in using adaptive methods as the basis for valid statistical inference is that selection bias can be difficult to quantify. In this paper, pairing the initial model with a complementary feature model enabled us to accomplish this goal in a simple yet principled way. In our simulation experiments, our method provides better error control while achieving nominal coverage rates in moderate sample sizes.

A number of important extensions and refinements are left open. Our current results only provide point-wise confidence intervals; extending our theory to the setting of multivariate testing or the setting of a generalized linear mixed models, seems like a promising avenue for further work. Another challenge is the selection of the proxy matrix $\boldsymbol{M}$ towards better efficiency and power. A systematic approach to design optimization and theory for such setting, would improve the finite sample performance. In general, work can be done to identify methods that furthermore allow for inference on the variance components and tests of heterogeneity even in more challenging circumstances, e.g., with small samples or a large number of covariates are likely to be bring impactful work to a broader scientific audience.

In this document we provide details to all of the technical results of the main document. In particular we provide detailed proofs of Theorems 1-4 and establish a sequence of useful Lemmas and their respective proofs.

# A Proofs of the main results

## A.1 Proof of Theorem 1

We observe that Theorem 1 is a direct consequence of the proofs of Theorems 3 and 4.

## A.2 Proof of Theorem 2

Theorem 2 is a consequence of Lemmas 5, 6 and 7 presented in Section B.

## A.3 Proof of Theorem 3 (under $H_0$)

Let us first notice that

$$\widehat{\sigma}_u \widehat{\sigma} T_n = \underbrace{n^{-1/2} \boldsymbol{U}^\top \widetilde{\boldsymbol{P}}(\boldsymbol{V} - \boldsymbol{X}\widehat{\gamma})}_{I_1} + \underbrace{n^{-1/2}(\boldsymbol{\theta}^* - \widehat{\boldsymbol{\theta}})^\top \boldsymbol{X}^\top \widetilde{\boldsymbol{P}}(\boldsymbol{V} - \boldsymbol{X}\widehat{\gamma})}_{I_2}. \tag{22}$$

By Theorem 1.6 of Rudelson and Zhou (2013), the restricted eigenvalue condition holds for $(\|\boldsymbol{\theta}^*\|_0, \kappa, \boldsymbol{X})$. Furthermore, by Lemma 5 we have $\mathbb{P}(\mathcal{A}) \to 1$ where

$$\mathcal{A} = \{\boldsymbol{\theta}^* \text{ and } \boldsymbol{\gamma}^* \text{are in the feasible regions of (12) and (10) }\}.$$

This, together with the restricted eigenvalue condition and Lemma 6 implies that with probability tending to 1

$$\left\|\widehat{\boldsymbol{\theta}} - \boldsymbol{\theta}^*\right\|_1 \le 8\eta_\theta \|\boldsymbol{\theta}^*\|_0 \kappa^{-2} \text{ and } n^{-1}\left\|\boldsymbol{X}(\widehat{\boldsymbol{\theta}} - \boldsymbol{\theta}^*)\right\|_2^2 \le 16\eta_\theta^2 \|\boldsymbol{\theta}^*\|_0 \kappa^{-2}. \tag{23}$$

Let us first bound $I_2$. We can write

$$|I_2| \le n^{1/2}\left\|\widehat{\boldsymbol{\theta}} - \boldsymbol{\theta}^*\right\|_1 \left\|n^{-1}\boldsymbol{X}^\top \widetilde{\boldsymbol{P}}(\boldsymbol{V} - \boldsymbol{X}\widehat{\gamma})\right\|_\infty.$$

Using (23) and the first constraint of the optimization problem (10), it holds on $\mathcal{M}$ that

$$|I_2| \le 8n^{1/2}\eta_\theta\eta_\gamma \|\boldsymbol{\theta}^*\|_0 \kappa^{-2}.$$

Let us now bound $|\widehat{\sigma}|^{-1} = (n^{-1}(\boldsymbol{V} - \boldsymbol{X}\widehat{\gamma})^\top \widetilde{\boldsymbol{P}}\widetilde{\boldsymbol{P}}(\boldsymbol{V} - \boldsymbol{X}\widehat{\gamma}))^{-1/2}$. By Lemma 3 used with $\boldsymbol{A} = \widetilde{\boldsymbol{P}}$, it holds that $|\widehat{\sigma}|^{-1} \le \bar{\eta}_\gamma^{-1}\sqrt{n^{-1}\boldsymbol{V}^\top \boldsymbol{V}}$ which is $\mathcal{O}_P(1)$ by Lemma 2. We thus have

$$\left|\frac{I_2}{\widehat{\sigma}}\right| \le 8n^{1/2}\eta_\theta\eta_\gamma \|\boldsymbol{\theta}^*\|_0 \kappa^{-2}\bar{\eta}_\gamma^{-1}\sqrt{n^{-1}\boldsymbol{V}^\top \boldsymbol{V}} = o_P(1) \tag{24}$$

where we used the sparsity condition $\|\boldsymbol{\theta}^*\|_0 = o(\sqrt{n}/\log(p)/\log(n))$ together with the rate of the tuning parameters.

Let us now consider $\frac{I_1}{\widehat{\sigma}}$. We write

$$\frac{I_1}{\widehat{\sigma}} = \boldsymbol{U}^\top \boldsymbol{D}$$

with $\boldsymbol{D} = n^{-1/2}\widehat{\sigma}^{-1}\widetilde{\boldsymbol{P}}(\boldsymbol{V} - \boldsymbol{X}\widehat{\boldsymbol{\gamma}})$ and observe that by construction, $\boldsymbol{D}^\top \boldsymbol{D} = 1$. Let us check that $\max_{1\leq i\leq n} |d_i u_i|$ and $\max_{1\leq i\leq n} d_i^2$ are both $o_P(1)$ in order to apply Lemma 9. First let us observe that

$$\max_{1\leq i\leq n} |d_i| = \mathcal{O}_P\left(\sqrt{\frac{\log n}{n}}\right).$$

Indeed, we have

$$\max_{1\leq i\leq n} |d_i| = n^{-1/2}\widehat{\sigma}^{-1}\left\|\widetilde{\boldsymbol{P}}(\boldsymbol{V} - \boldsymbol{X}\widehat{\boldsymbol{\gamma}})\right\|_\infty \overset{(i)}{\leq} n^{-1/2}\bar{\eta}_\gamma^{-1}\sqrt{n^{-1}\boldsymbol{V}^\top\boldsymbol{V}}\mu_\gamma \overset{(ii)}{=} \mathcal{O}_P\left(\sqrt{\frac{\log n}{n}}\right)$$

where $(i)$ follows from Lemma 3 and from the third constraint of (10) and $(ii)$ follows from Lemma 2 and $\mu_\gamma \asymp \sqrt{\log n}$.

We also observe that $\max_{1\leq i\leq n} |u_i| = \mathcal{O}_P(\sqrt{\log n})$ by the union bound and the sub-Gaussian property. We thus have

$$\max_{1\leq i\leq n} |d_i u_i| \leq \max_{1\leq i\leq n} |d_i| \max_{1\leq i\leq n} |u_i| = \mathcal{O}_P(n^{-1/2}\log n) = o_P(1)$$

and $\max_{1\leq i\leq n} w_i^2 = \mathcal{O}_P(n^{-1}\log n) = o_P(1)$. It follows by Lemma 9 applied on $\{u_i\}_{i=1}^n$ and $\{d_i\}_{i=1}^n$ that

$$\frac{I_1}{\widehat{\sigma}} \xrightarrow{d} \mathcal{N}(0, \sigma_u^2) \tag{25}$$

By combining (22), (24) and (25) we have $\widehat{\sigma}_u T_n \xrightarrow{d} \mathcal{N}(0, \sigma_u^2)$.

Since $n^{-1/2}\left\|\boldsymbol{X}(\boldsymbol{\theta}^* - \widehat{\boldsymbol{\theta}})\right\|_2 = o_P(1)$, we have

$$\widehat{\sigma}_u = n^{-1/2}\left\|\boldsymbol{Z} - \boldsymbol{X}\widehat{\boldsymbol{\theta}}\right\|_2 = n^{-1/2}\left\|\boldsymbol{U}\right\|_2 + o_P(1) = \sigma_u + o_P(1).$$

Using Slutzky's lemma, we obtain that $T_n \xrightarrow{d} \mathcal{N}(0,1)$ which ends the proof of Theorem 3. □

### A.4 Proof of Theorem 4 (under $H_1$)

First, note that by Lemma 5 we have $\mathbb{P}(\mathcal{A}) \to 1$ where

$$\mathcal{A} = \{\theta^* \text{ and } \gamma^* \text{are in the feasible regions of (12) and (10) }\}.$$

Furthermore, by Theorem 1.6 of Rudelson and Zhou (2013) and since $\widetilde{\boldsymbol{P}}$ satisfies the $P$-condition, the restricted eigenvalue condition holds for $(\|\boldsymbol{\theta}^*\|_0, \kappa, \boldsymbol{X})$ and for $(\|\boldsymbol{\gamma}^*\|_0, \kappa, \widetilde{\boldsymbol{P}}^{1/2}\boldsymbol{X})$ with probability tending to 1. It follows by Lemma 6 applied to problems (12) and (10) that we probability tending to 1

$$\begin{aligned}
\left\|\widehat{\boldsymbol{\theta}} - \boldsymbol{\theta}^*\right\|_1 &\leq 8\eta_\theta \|\boldsymbol{\theta}^*\|_0 \kappa^{-2} \quad \text{and} \quad n^{-1}\left\|\boldsymbol{X}(\widehat{\boldsymbol{\theta}} - \boldsymbol{\theta}^*)\right\|_2^2 \leq 16\eta_\theta^2 \|\boldsymbol{\theta}^*\|_0 \kappa^{-2} \\
\|\widehat{\boldsymbol{\gamma}} - \boldsymbol{\gamma}^*\|_1 &\leq 8\eta_\gamma \|\boldsymbol{\gamma}^*\|_0 \kappa^{-2} \quad \text{and} \quad n^{-1}\left\|\widetilde{\boldsymbol{P}}^{1/2}\boldsymbol{X}(\widehat{\boldsymbol{\gamma}} - \boldsymbol{\gamma}^*)\right\|_2^2 \leq 16\eta_\gamma^2 \|\boldsymbol{\gamma}^*\|_0 \kappa^{-2}.
\end{aligned} \tag{26}$$

Since $\boldsymbol{V} = \boldsymbol{X}\boldsymbol{\gamma}^* + \boldsymbol{Z}hn^{-1/2} + \boldsymbol{W}\boldsymbol{b} + \boldsymbol{\epsilon}$, we can write the following decomposition:

$$\widehat{\sigma}_u \widehat{\sigma} T_n = \underbrace{n^{-1/2}(\boldsymbol{W}\boldsymbol{b} + \boldsymbol{\epsilon})^\top \widetilde{\boldsymbol{P}}(\boldsymbol{Z} - \boldsymbol{X}\widehat{\boldsymbol{\theta}})}_{T_1} + \underbrace{n^{-1/2}(\boldsymbol{\gamma}^* - \widehat{\boldsymbol{\gamma}})^\top \boldsymbol{X}^\top \widetilde{\boldsymbol{P}}(\boldsymbol{Z} - \boldsymbol{X}\widehat{\boldsymbol{\theta}})}_{T_2} + \underbrace{n^{-1}h\boldsymbol{Z}^\top \widetilde{\boldsymbol{P}}(\boldsymbol{Z} - \boldsymbol{X}\widehat{\boldsymbol{\theta}})}_{T_3}. \tag{27}$$

Let us first consider $T_2$. We observe that

$$|T_2| \leq n^{1/2} \|\widehat{\boldsymbol{\gamma}} - \boldsymbol{\gamma}^*\|_1 \left\| n^{-1}\boldsymbol{X}^\top \widetilde{\boldsymbol{P}}(\boldsymbol{Z} - \boldsymbol{X}\widehat{\boldsymbol{\theta}}) \right\|_\infty \leq n^{1/2} 8\eta_\gamma \|\boldsymbol{\gamma}^*\|_0 \kappa^{-2}\eta'_\theta = o_P(1) \tag{28}$$

where we used (26) and the fourth constraint of (12).

Regarding $T_3$, we write

$$T_3 = n^{-1}h\boldsymbol{Z}^\top \widetilde{\boldsymbol{P}}\boldsymbol{U} + n^{-1}h\boldsymbol{Z}\widetilde{\boldsymbol{P}}\boldsymbol{X}(\boldsymbol{\theta}^* - \widehat{\boldsymbol{\theta}}) := T_{31} + T_{32}.$$

For $T_{32}$, we observe that $\widetilde{\boldsymbol{P}}\boldsymbol{Z}$ is sub-Gaussian since $\boldsymbol{Z}$ is sub-Gaussian and $\widetilde{\boldsymbol{P}}$ satisfies the $P$-condition. This together with (26) implies on $\mathcal{M}$ that

$$|T_{32}| \leq h \left\| \boldsymbol{n}^{-1/2}\widetilde{\boldsymbol{P}}\boldsymbol{Z} \right\|_2 \left\| \boldsymbol{n}^{-1/2}\boldsymbol{X}(\boldsymbol{\theta}^* - \widehat{\boldsymbol{\theta}}) \right\|_2 = \mathcal{O}_P(\eta_\theta \sqrt{\|\boldsymbol{\theta}^*\|_0}) = o_P(1).$$

For $T_{31}$, we use the Markov law of large numbers for independent but non-identical random variables. Working on a group level, we write

$$T_{31} = hn^{-1}\sum_{j=1}^{N} \boldsymbol{Z}_j \widetilde{\boldsymbol{P}}_j \boldsymbol{U}_j = hn^{-1}\sum_{j=1}^{N} d_j$$

with $\widetilde{\boldsymbol{P}} = \text{diag}(\widetilde{\boldsymbol{P}}_1, \ldots, \widetilde{\boldsymbol{P}}_N)$ and $\widetilde{\boldsymbol{P}}_j \in \mathbb{R}^{n_j \times n_j}$ and we observe that $d_j \perp d_{j'}$ for $g \neq j'$. Let us show that the mean and the variance of $d_j$ are bounded. We compute

$$\mathbb{E}[d_j] = \mathbb{E}[\boldsymbol{U}_j^\top \widetilde{\boldsymbol{P}}_j \boldsymbol{U}_j] + \mathbb{E}[\boldsymbol{\theta}^{*\top} \boldsymbol{X}_j^\top \widetilde{\boldsymbol{P}}_j \boldsymbol{U}_j] = \sigma_u^2 \text{trace}(\widetilde{\boldsymbol{P}}_j)$$

where we make use of the independence of $\boldsymbol{X}$ and $\boldsymbol{U}$. For the variance term, we make use of the variance formula of a quadratic form (see appendix) and obtain

$$\text{Var}[\boldsymbol{U}_j^\top \widetilde{\boldsymbol{P}}_j \boldsymbol{U}_j] = 2\sigma_u^4 \text{trace}(\widetilde{\boldsymbol{P}}_j \widetilde{\boldsymbol{P}}_j).$$

This implies that the variance of $d_j$ is bounded. By using the Markov's law of large numbers on $\{d_j\}_{j=1}^N$ we obtain that

$$N^{-1}\sum_{j=1}^{N}(d_j - \sigma_u^2 \text{trace}(\widetilde{\boldsymbol{P}}_j)) = o_P(1).$$

Multiplying this result by the bounded quantity $h\frac{N}{n}$, we get $T_{31} - hn^{-1}\sigma_u^2 \text{trace}(\widetilde{\boldsymbol{P}}) = o_P(1)$. We thus have

$$T_3 = h\sigma_u^2 n^{-1} \text{trace}(\widetilde{\boldsymbol{P}}) + o_P(1). \tag{29}$$

Let us now prove that $T_1$ converges to a normal distribution under the alternative hypothesis. We write

$$T_1/\widehat{\sigma}_u = \boldsymbol{r}^\top \widetilde{\boldsymbol{P}}(\boldsymbol{W}\boldsymbol{b} + \boldsymbol{\epsilon})$$

with $\boldsymbol{r} = n^{-1/2}\widehat{\sigma}_u^{-1}(\boldsymbol{Z} - \boldsymbol{X}\widehat{\boldsymbol{\theta}})$. Let us define $\boldsymbol{\epsilon}^* = \widetilde{\boldsymbol{P}}(\boldsymbol{W}\boldsymbol{b} + \boldsymbol{\epsilon})$ and the sequences of random vectors $\{\boldsymbol{r}_j\}_{j=1}^N$ and $\left\{\boldsymbol{\epsilon}_j^*\right\}_{j=1}^N$ induced by the group structure of the mixed model. Let us show that the conditions of Lemma 10 are satisfied for these sequences of random vectors.

First, we see that $\boldsymbol{\epsilon}_j^* \in \mathbb{R}^{n_j}$ are mean zero, sub-Gaussian random vectors (as linear combination of sub-Gaussian random variables) and such that $\text{var}(\boldsymbol{\epsilon}_j^*) = \boldsymbol{K}_j$ with $\boldsymbol{K}_j$ being a block diagonal element of $\sigma_\epsilon^2\widetilde{\boldsymbol{P}}\boldsymbol{P}^{-1}\widetilde{\boldsymbol{P}}$. By construction, the elements of $\boldsymbol{K}_j$ are bounded which implies that $\lambda_{\max}(\boldsymbol{K}_j)$ is bounded in $N$. Second, we notice that by definition of $\widehat{\sigma}_u$, we have that $\sum_{j=1}^N \boldsymbol{r}_j^\top \boldsymbol{r}_j = \boldsymbol{r}^\top\boldsymbol{r} = 1$.

As proven in Theorem 3, it holds that $\widehat{\sigma}_u = \sigma_u + o_P(1)$. By the continuous mapping theorem, we deduce that $\widehat{\sigma}_u^{-1} = \sigma_u + o_P(1) = \mathcal{O}_P(1)$. Thus, by the third constraint of (12) we have that

$$\|\boldsymbol{r}\|_\infty \leq n^{-1/2}\mu_\theta/\widehat{\sigma}_u = \mathcal{O}_P(n^{-1/2}\mu_\theta) = o_P(1).$$

This implies that $\max_j \boldsymbol{r}_j^\top\boldsymbol{r}_j \leq \max_j n_j \|\boldsymbol{r}\|_\infty^2 = o_P(1)$ since the $n_j$ are bounded. Furthermore, we observe that since $\boldsymbol{\epsilon}^*$ is sub-Gaussian, it holds by the union bound that $\|\boldsymbol{\epsilon}^*\|_\infty = \mathcal{O}_P(\sqrt{\log n})$. We thus have

$$\max_{1\leq j\leq n}\left|\boldsymbol{r}_j^\top\boldsymbol{\epsilon}_j^*\right| \leq \|\ r\|_\infty \|\boldsymbol{\epsilon}^*\|_\infty \, \mathcal{O}_P(n^{-1/2}\mu_\theta\sqrt{\log n}) = o_P(1).$$

We can now apply Lemma 10 on $\{\boldsymbol{r}_j\}_{j=1}^N$ and $\left\{\boldsymbol{\epsilon}_j^*\right\}_{j=1}^N$ and derive that

$$\frac{\widehat{\sigma}_u^{-1}T_1}{\sqrt{\sum_{j=1}^N \boldsymbol{r}_j^\top\boldsymbol{K}_j\boldsymbol{r}_j}} \xrightarrow{d} \mathcal{N}(0,1). \tag{30}$$

By Lemmas 7 and 8, it holds that $\widehat{\sigma} = \sqrt{\sum_{j=1}^N \boldsymbol{r}_j^\top\boldsymbol{K}_j\boldsymbol{r}_j} + o_P(1)$. This, together with (30) and Slutsky's Theorem implies that

$$\widehat{\sigma}_u^{-1}\widehat{\sigma}^{-1}T_1 \xrightarrow{d} \mathcal{N}(0,1). \tag{31}$$

Now, using $\widehat{\sigma}_u = \sigma_u + o_P(1)$ we can combine (28), (29) and (31) and obtain that

$$T_n = (\widehat{\sigma}^{-1}\widehat{\sigma}_u^{-1})(T_1 + T_2 + T_3) \stackrel{d}{=} \mathcal{N}(0,1) + h\sigma_u\sigma_\epsilon^{-1}\frac{n^{-1}\text{trace}(\widetilde{\boldsymbol{P}})}{\sqrt{\text{trace}(n^{-1}\widetilde{\boldsymbol{P}}\boldsymbol{P}^{-1}\widetilde{\boldsymbol{P}})}} + o_P(1). \tag{32}$$

This ends the proof of Theorem 4. □

# B Auxiliary proofs and statements

**Lemma 1.** *The matrix* $\boldsymbol{P} = (\boldsymbol{I}_n - \boldsymbol{W}\boldsymbol{E}^{-1}\boldsymbol{W}^\top)(\boldsymbol{I}_n - \boldsymbol{W}\boldsymbol{E}^{-1}\boldsymbol{W}^\top) + \sigma_\epsilon^2\boldsymbol{W}\boldsymbol{E}^{-1}\boldsymbol{\Psi}^{-1}\boldsymbol{E}^{-1}\boldsymbol{W}^\top$, *with* $\boldsymbol{E} = \boldsymbol{W}^\top\boldsymbol{W} + \sigma_\epsilon^2\boldsymbol{\Psi}^{-1}$, *can be rewritten as* $\boldsymbol{P} = (\boldsymbol{I}_n + \sigma_\epsilon^{-2}\boldsymbol{W}\boldsymbol{\Psi}\boldsymbol{W}^\top)^{-1}$.

**Proof of Lemma 1.** A proof can be found in the supplementary material of Fan and Li (2012).

**Lemma 2.** *Under* $H_0 : \beta^* = \beta_0$ *and under conditions of Theorem 3, it holds that* $n^{-1}\boldsymbol{V}^\top\boldsymbol{V} = \mathcal{O}_P(1)$.

**Proof of Lemma 2.** Under $H_0$ we have $\boldsymbol{V} = \boldsymbol{X}\boldsymbol{\gamma}^* + \boldsymbol{\epsilon}^*$ with $\boldsymbol{\epsilon}^* = \boldsymbol{\epsilon} + \boldsymbol{W}\boldsymbol{b}$. Recall that $\text{var}(\boldsymbol{\epsilon}^*) = \boldsymbol{P}^{-1}$. Let us work at a group level and consider the sequence of random variables $\left\{\boldsymbol{V}_j^\top\boldsymbol{V}_j\right\}_{j=1}^N$. We

have

$$\begin{aligned}\mathbb{E}[\boldsymbol{V}_j^\top \boldsymbol{V}_j] &= \mathbb{E}[\boldsymbol{\epsilon}_j^{*\top}\boldsymbol{\epsilon}_j^*] + \mathbb{E}[\boldsymbol{\gamma}^{*\top}\boldsymbol{X}_j^\top \boldsymbol{X}_j\boldsymbol{\gamma}^*] + \mathbb{E}[\boldsymbol{\epsilon}_j^{*\top}\boldsymbol{X}_j\boldsymbol{\gamma}^*]\\ &= \text{trace}(\boldsymbol{P}_j^{-1}) + n_j\boldsymbol{\gamma}^{*\top}\boldsymbol{\Sigma}\boldsymbol{\gamma}^* + 0\\ &= \mathcal{O}(1)\end{aligned}$$

by condition 1 and by the boundedness of the elements of $\boldsymbol{P}^{-1}$. With similar arguments, we can show that $\text{Var}[\boldsymbol{V}_j^\top \boldsymbol{V}_j]$ is bounded. It follows from the Markov's law of large number that

$$N^{-1}\sum_{j=1}^{N}(\boldsymbol{V}_j^\top \boldsymbol{V}_j - \mathbb{E}[\boldsymbol{V}_j^\top \boldsymbol{V}_j]) \xrightarrow{p} 0.$$

which implies that $n^{-1}\boldsymbol{V}^\top\boldsymbol{V} = \mathcal{O}_P(1)$. □

**Lemma 3.** *Let $\boldsymbol{A}$ be a $n\times n$ symmetric matrix. If $n^{-1}\boldsymbol{V}^\top\boldsymbol{A}(\boldsymbol{V}-\boldsymbol{X}\widehat{\gamma}) \geq \bar{\eta}$, then $n^{-1}(\boldsymbol{V}-\boldsymbol{X}\widehat{\gamma})^\top\boldsymbol{A}\boldsymbol{A}(\boldsymbol{V}-\boldsymbol{X}\widehat{\gamma}) \geq \bar{\eta}^2/(n^{-1}\boldsymbol{V}^\top\boldsymbol{V})$.*

**Proof of Lemma 3.**
For any $a>0$, we have

$$\begin{aligned}&n^{-1}(\boldsymbol{V}-\boldsymbol{X}\widehat{\gamma})^\top\boldsymbol{A}\boldsymbol{A}(\boldsymbol{V}-\boldsymbol{X}\widehat{\gamma})\\ &\quad\geq\ n^{-1}(\boldsymbol{V}-\boldsymbol{X}\widehat{\gamma})^\top\boldsymbol{A}\boldsymbol{A}(\boldsymbol{V}-\boldsymbol{X}\widehat{\gamma}) + a(\bar{\eta} - n^{-1}\boldsymbol{V}^\top\boldsymbol{A}(\boldsymbol{V}-\boldsymbol{X}\widehat{\gamma}))\\ &\quad\geq\ \min_{\gamma}\left\{n^{-1}(\boldsymbol{V}-\boldsymbol{X}\gamma)^\top\boldsymbol{A}\boldsymbol{A}(\boldsymbol{V}-\boldsymbol{X}\gamma) + a(\bar{\eta} - n^{-1}\boldsymbol{V}^\top\boldsymbol{A}(\boldsymbol{V}-\boldsymbol{X}\gamma))\right\}\\ &\quad=\ a\bar{\eta} - \frac{1}{4}a^2n^{-1}\boldsymbol{V}^\top\boldsymbol{V}\end{aligned}$$

where the last equality follows from the first-order condition of the quadratic problem which is

$$n^{-1}\boldsymbol{X}^\top\boldsymbol{A}\left\{\frac{1}{2}a\boldsymbol{V} - \boldsymbol{A}(\boldsymbol{V}-\boldsymbol{X}\gamma)\right\} = 0.$$

Then by choosing $a = 2\bar{\eta}/(n^{-1}\boldsymbol{V}^\top\boldsymbol{V})$ we obtain

$$n^{-1}(\boldsymbol{V}-\boldsymbol{X}\widehat{\gamma})^\top\boldsymbol{A}\boldsymbol{A}(\boldsymbol{V}-\boldsymbol{X}\widehat{\gamma}) \geq \bar{\eta}^2/(n^{-1}\boldsymbol{V}^\top\boldsymbol{V}). \quad \square$$

**Lemma 4.** *The matrices $\boldsymbol{P}$ and $\boldsymbol{P}^{-1}$ satisfy the P-condition where $\boldsymbol{P} = (\boldsymbol{I}_n + \sigma_\epsilon^{-2}\boldsymbol{W}\boldsymbol{\Psi}\boldsymbol{W}^\top)^{-1}$. Furthermore, for any $Nq\times Nq$ matrix $\boldsymbol{M}$ having the same block diagonal structure as $\boldsymbol{W}^\top\boldsymbol{W}$, it holds that the matrix $\widetilde{\boldsymbol{P}} = (\boldsymbol{I}_n + \boldsymbol{W}\boldsymbol{M}\boldsymbol{W}^\top)^{-1}$, if it exists, satisfies the P-condition.*

**Proof of Lemma 4.** First, note that by construction $\boldsymbol{P}$, $\boldsymbol{P}^{-1}$ and $\widetilde{\boldsymbol{P}}$ have the correct block diagonal structure. Furthermore, by condition 1 it is clear that the elements of $\boldsymbol{P}^{-1}$ are bounded. Let us now show that each element of $\boldsymbol{P}$ and $\widetilde{\boldsymbol{P}}$ is bounded. Recall that the $\ell_2$ norm of a matrix $\boldsymbol{A}$ is defined as $\|\boldsymbol{A}\|_2 = \sup_{x\neq 0}\frac{\|\boldsymbol{A}x\|_2}{\|\boldsymbol{x}\|_2} = \sigma_{\max}(\boldsymbol{A})$ where $\sigma_{\max}(\boldsymbol{A})$ is the largest singular value of $\boldsymbol{A}$. By consequence, it suffices to show that $\sigma_{\max}(\widetilde{\boldsymbol{P}})$ and $\sigma_{\max}(\boldsymbol{P})$ are bounded to guarantee that each element is bounded.

$$\begin{aligned}
\sigma_{\max}(\widetilde{\boldsymbol{P}}) &= \max\left\{\sigma_1(\widetilde{\boldsymbol{P}}),\ldots,\sigma_n(\widetilde{\boldsymbol{P}})\right\} \\
&= \max\left\{\sigma_1(\widetilde{\boldsymbol{P}}^{-1})^{-1},\ldots,\sigma_n(\widetilde{\boldsymbol{P}}^{-1})^{-1}\right\} \\
&= \max\left\{(1+\sigma_1(\boldsymbol{W}\boldsymbol{M}\boldsymbol{W}^\top))^{-1},\ldots,(1+\sigma_n(\boldsymbol{W}\boldsymbol{M}\boldsymbol{W}^\top))^{-1}\right\} \\
&\leq 1
\end{aligned}$$

since by definition a singular value is always non-negative. The proof for $\boldsymbol{P}$ and $\boldsymbol{P}^{-1}$ follow from similar arguments.

□

The following Lemma, combined with Lemma 4 guarantees the feasibility of the optimization problems (12) and (10), which is required to derive the properties of our estimators.

**Lemma 5.** *Let $\boldsymbol{A} \in \mathbb{R}^{n\times n}$ satisfying the $P$-condition. Suppose that condition 1 holds and that $\beta^* = \beta_0 + hn^{-1/2}$ for a constant $h \in \mathbb{R}$. Then there exist constants $C_1$, $C_2$, $c_0 > 0$ such that for any $\eta_\gamma, \eta_\theta, \eta'_\theta \geq C_1 \log n\sqrt{n^{-1}\log p}$, $\mu_\gamma, \mu_\theta \geq C_2\sqrt{\log n}$, $\bar{\eta}_\gamma, \in \left(0, n^{-1}\text{trace}(\sigma_\epsilon^2 \boldsymbol{A}\boldsymbol{P}^{-1}) - c_0\right)$ and $\bar{\eta}_\theta, \in \left(0, \sigma_u^2 - c_0\right)$, it holds with probability tending to 1 that*

$$\begin{aligned}
\left\|n^{-1}\boldsymbol{X}^\top\boldsymbol{A}(\boldsymbol{V}-\boldsymbol{X}\boldsymbol{\gamma}^*)\right\|_\infty &\leq \eta_\gamma \\
n^{-1}\boldsymbol{V}^\top\boldsymbol{A}(\boldsymbol{V}-\boldsymbol{X}\boldsymbol{\gamma}^*) &\geq \bar{\eta}_\gamma \\
\left\|\boldsymbol{A}(\boldsymbol{V}-\boldsymbol{X}\boldsymbol{\gamma}^*)\right\|_\infty &\leq \mu_\gamma
\end{aligned} \tag{33}$$

*and*

$$\begin{aligned}
\left\|n^{-1}\boldsymbol{X}^\top(\boldsymbol{Z}-\boldsymbol{X}\boldsymbol{\theta})\right\|_\infty &\leq \eta_\theta \\
n^{-1}\boldsymbol{Z}^\top(\boldsymbol{Z}-\boldsymbol{X}\boldsymbol{\theta}) &\geq \bar{\eta}_\theta \\
\left\|\boldsymbol{Z}-\boldsymbol{X}\boldsymbol{\theta}\right\|_\infty &\leq \mu_\theta \\
\left\|n^{-1}\boldsymbol{X}^\top\boldsymbol{A}(\boldsymbol{Z}-\boldsymbol{X}\boldsymbol{\theta})\right\|_\infty &\leq \eta'_\theta
\end{aligned} \tag{34}$$

**Proof of Lemma 5.** We prove that (33) holds with probability tending to 1. The proof for (34) follows from similar arguments and is thus omitted.

Let us first show that there exists a constant $C_1$ such that with probability converging to 1 it holds that $\left\|n^{-1}\boldsymbol{X}^\top\boldsymbol{A}(\boldsymbol{V}-\boldsymbol{X}\boldsymbol{\gamma}^*)\right\|_\infty \leq C_1\sqrt{n^{-1}\log p}$. Using the block diagonal structure of $\boldsymbol{A} = (\boldsymbol{A}_1,\ldots,\boldsymbol{A}_N)$ associated with the group structure, we observe that

$$n^{-1}\boldsymbol{X}^\top\boldsymbol{A}(\boldsymbol{V}-\boldsymbol{X}\boldsymbol{\gamma}^*) = \frac{1}{n}\sum_{g=1}^N \boldsymbol{X}_{(g)}^\top\boldsymbol{A}_g(\boldsymbol{V}_{(g)} - \boldsymbol{X}_{(g)}\boldsymbol{\gamma}^*) = \frac{1}{n}\sum_{g=1}^N \boldsymbol{d}_g \tag{35}$$

where the subscript $(g)$ refers to the fact that we extract the rows corresponding to goup $g$.

**Claim 1:** Each component of $\boldsymbol{d}_g \in \mathbb{R}^p$ has an exponential-type tail with parameters $(K_1, \frac{1}{2})$.

To show this, let us decompose the $j$-th component as follows:

$$d_g^j = \sum_{i\in\text{group}g}\sum_{k\in\text{group}g} x_{ij}a_{ik}(v_k - \boldsymbol{X}_k^\top\boldsymbol{\gamma}^*). \tag{36}$$

As the group $g$ contains a bounded number of elements $n_g$, it suffices to show that each $x_{ij}a_{ik}(v_k - \boldsymbol{X}_k^\top\boldsymbol{\gamma}^*)$ has an exponential-type tail with parameters $(K_0, \frac{1}{2})$ to prove the claim.

We note that

$$v_k - \boldsymbol{X}_k^\top\boldsymbol{\gamma}^* = z_k hn^{-1/2} + \epsilon_k + w_k\boldsymbol{b}_g$$

with $\mathrm{var}(\boldsymbol{b}_g) = \psi \in \mathbb{R}^{q\times q}$. The sub-Gaussianity of $\epsilon_k$ and of each component of $\boldsymbol{b}_g$ and the fact that each component of $w_k$ is bounded imply by the properties of sub-Gaussian random variables that $v_k - \boldsymbol{X}_k^\top \boldsymbol{\gamma}^*$ is sub-Gaussian. Now, using that $x_{ij}$ is sub-Gaussian and that $a_{ij}$ is $\mathcal{O}(1)$ we obtain by Lemma 5.14 of Vershynin (2010) that $x_{ij}a_{ik}(v_k - \boldsymbol{X}_k^\top \boldsymbol{\gamma}^*)$ has an exponential-type tail with parameters $(K_0, \frac{1}{2})$. This concludes the proof of claim 1.

Observe that $d_g^j \perp d_{g'}^j$. Thanks to the exponential-type tail property of $\left\{d_g^j\right\}_{g=1}^N$, it follows from Proposition 5.16 of Vershynin (2010) that there exist constants $K_2$ and $K_3$ such that for any $x > 0$ we have

$$\mathbb{P}\left(N^{-1/2}\left|\sum_{g=1}^N (d_g^j - \mathbb{E}d_g^j)\right| > x\right) \leq K_2 \exp(-K_3 x^2). \tag{37}$$

We compute

$$\begin{aligned}
\mathbb{E}d_g^j &= \mathbb{E}\left[\boldsymbol{X}_{(g)j}^\top \boldsymbol{A}_g(n^{-1/2}h\boldsymbol{Z}_{(g)} + \boldsymbol{\epsilon}_{(g)} + \boldsymbol{W}_{(g)}\boldsymbol{b}_g)\right] \\
&\overset{(i)}{=} \mathbb{E}\left[\boldsymbol{X}_{(g)j}^\top \boldsymbol{A}_g n^{-1/2} h \boldsymbol{X}_{(g)}\boldsymbol{\theta}^*\right] \\
&\overset{(ii)}{=} hn^{-1/2}\mathrm{trace}(\boldsymbol{A}_g)\boldsymbol{\Sigma}_{j:}\boldsymbol{\theta}^*
\end{aligned}$$

where *(i)* comes from the independence of $\boldsymbol{X}_{(g)j}$ with $\boldsymbol{u}$, $\boldsymbol{\epsilon}$ and $\boldsymbol{W}\boldsymbol{b}$ and *(ii)* comes from the independence between $x_{kj}$ and $x_{lj}$ for $k \neq l$, this is,

$$\mathbb{E}\left[\boldsymbol{X}_{(g)j}^\top \boldsymbol{A}_g \boldsymbol{X}_{(g)})\right] = \sum_{k\in\mathrm{group}g}\sum_{l\in\mathrm{group}g} \mathbb{E}\left[x_{kj}A_{kl}x_{lj}\right] = \sum_{k\in\mathrm{group}g} \mathbb{E}\left[x_{kj}A_{kk}x_{kj}\right] = \mathrm{trace}(\boldsymbol{A}_g)\boldsymbol{\Sigma}_{j:}.$$

Thus we obtain that $\frac{1}{N}\sum_{g=1}^N \mathbb{E}[d_g^j] = hn^{-1/2}N^{-1}\mathrm{trace}(\boldsymbol{A})\Sigma_{j:}$. Now, observe that since $\boldsymbol{A}$ satisfies the $P$-condition it holds that $n^{-1}\mathrm{trace}(\boldsymbol{A})$ is bounded. It follows that there exists a constant $K_4$ such that

$$\left|\frac{1}{N}\sum_{g=1}^N \mathbb{E}[d_g^j]\right| \leq K_4 \mathrm{trace}(n^{-1}\boldsymbol{A})\frac{n}{N}\sqrt{n^{-1}\log p} \tag{38}$$

Combining (35), (37) and (38) and using the union bound property and the inequality $(a-b)^2 \geq$

$\frac{a^2}{2} - b^2$, we derive the following result:

$$
\begin{aligned}
& \mathbb{P}\left(\left\|n^{-1}\boldsymbol{X}^\top \boldsymbol{A}(\boldsymbol{V}-\boldsymbol{X}\boldsymbol{\gamma}^*)\right\|_\infty > \eta_\gamma\right) \\
\leq \;& p\,\mathbb{P}\left(\left|n^{-1}\boldsymbol{X}_j^\top \boldsymbol{A}(\boldsymbol{V}-\boldsymbol{X}\boldsymbol{\gamma}^*)\right| > \eta_\gamma\right) \\
= \;& p\,\mathbb{P}\left(\left|N^{-1/2}\sum_{g=1}^N d_g^j\right| > \eta_\gamma \frac{n}{\sqrt{N}}\right) \\
= \;& p\,\mathbb{P}\left(\left|N^{-1/2}\sum_{g=1}^N (d_g^j - \mathbb{E}[d_g^j])\right| > \eta_\gamma \frac{n}{\sqrt{N}} - K_4 \text{trace}(n^{-1}\boldsymbol{A})\frac{n}{\sqrt{N}}\sqrt{n^{-1}\log p}\right) \\
\leq \;& K_2 p\,\exp\left[-K_3\left(\eta_\gamma \frac{n}{\sqrt{N}} - K_4 \text{trace}(n^{-1}\boldsymbol{A})\frac{n}{\sqrt{N}}\sqrt{n^{-1}\log p}\right)^2\right] \\
\leq \;& K_2 p\,\exp\left[-K_3 n\left(\frac{1}{2}\eta_\gamma^2 \frac{n}{N} - K_4^2 \text{trace}(n^{-1}\boldsymbol{A})^2\frac{n}{N} n^{-1}\log p\right)\right] \\
= \;& K_2 p^{1+K_3K_4^2\text{trace}(n^{-1}\boldsymbol{A})^2\frac{n}{N}}\,\exp\left(-\frac{K_3}{2} n\eta_\gamma^2 \frac{n}{N}\right).
\end{aligned}
$$

By definition of $\boldsymbol{A}$, $\text{trace}(n^{-1}\boldsymbol{A}) \leq \log^2 n$. This implies that for a constant $C_1$ large enough and for $\eta_\gamma \geq C_1(\log n)\sqrt{n^{-1}\log p}$ we have $\mathbb{P}\left(\left\|n^{-1}\boldsymbol{X}^\top \boldsymbol{A}(\boldsymbol{V}-\boldsymbol{X}\boldsymbol{\gamma}^*)\right\|_\infty \leq \eta_\gamma\right)$ with probability tending to 1.

This ends the first part of the proof.

Let us now tackle the third constraint of (33) and show that there exists a constant $C_2 > 0$ such that for any $\mu_\gamma > C_2\sqrt{\log n}$, we have $\|\boldsymbol{A}(\boldsymbol{V}-\boldsymbol{X}\boldsymbol{\gamma}^*)\|_\infty \leq C_2\sqrt{\log n}$ with probability tending to 1. We first observe that each component of $\boldsymbol{A}(\boldsymbol{V}-\boldsymbol{X}\boldsymbol{\gamma}^*)$ is sub-Gaussian because of the structure of $\boldsymbol{A}$ and because components of $\boldsymbol{V}-\boldsymbol{X}\boldsymbol{\gamma}^*$ are sub-Gaussian. By the union bound and the definition of a sub-Gaussian variable, there exists a constant $K_4 > 0$ such that for any $x > 0$

$$
\mathbb{P}\left(\|\boldsymbol{A}(\boldsymbol{V}-\boldsymbol{X}\boldsymbol{\gamma}^*)\|_\infty > x\sqrt{\log n}\right) \leq n\exp(1 - K_5 x^2 \log n). \tag{39}
$$

The result follows by taking a constant $x$ large enough. To conclude, let us tackle the second constraint of (33). Let us show that for any $\bar{\eta}_\gamma \in \left(0, \text{trace}(n^{-1}\sigma_\epsilon^2 \boldsymbol{A}\boldsymbol{P}^{-1}) - c_0\right)$, we have

$$
\mathbb{P}\left(n^{-1}\boldsymbol{V}^\top \boldsymbol{A}(\boldsymbol{V}-\boldsymbol{X}\boldsymbol{\gamma}^*) < \bar{\eta}_\gamma\right) \to 0. \tag{40}
$$

Working at the group level induced by the block-diagonal structure of $\boldsymbol{A}$, we write

$$
n^{-1}\boldsymbol{V}^\top \boldsymbol{A}(\boldsymbol{V}-\boldsymbol{X}\boldsymbol{\gamma}^*) = n^{-1}\sum_{j=1}^N \boldsymbol{V}_j \boldsymbol{A}_j(\boldsymbol{V}_j - \boldsymbol{X}_j\boldsymbol{\gamma}^*) := n^{-1}\sum_{j=1}^N b_j
$$

Let us consider the following decomposition:

$$
b_j = \underbrace{(\boldsymbol{V}_j - \boldsymbol{X}_j\boldsymbol{\gamma}^*)^\top \boldsymbol{A}_j(\boldsymbol{V}_j - \boldsymbol{X}_j\boldsymbol{\gamma}^*)}_{b_{j,1}} + \underbrace{\boldsymbol{\gamma}^{*\top}\boldsymbol{X}_j^\top \boldsymbol{A}_j(\boldsymbol{V}_j - \boldsymbol{X}_j\boldsymbol{\gamma}^*)}_{b_{j,2}}.
$$

Using the equality

$$
\boldsymbol{V}_j - \boldsymbol{X}_j\boldsymbol{\gamma}^* = \boldsymbol{W}_j\boldsymbol{b}_j + \boldsymbol{\epsilon}_j + hn^{-1/2}(\boldsymbol{X}_j\boldsymbol{\theta}^* + \boldsymbol{u}_j)
$$

and the independence between $\boldsymbol{X}_j$ and $\boldsymbol{\epsilon}_j$, $\boldsymbol{b}_j$ and $\boldsymbol{u}_j$, we obtain that

$$\mathbb{E}[b_{j,2}] = hn^{-1/2}\boldsymbol{\gamma}^{*\top}\mathbf{E}[\boldsymbol{X}_j^\top \boldsymbol{A}_j \boldsymbol{X}_j]\boldsymbol{\theta}^* \tag{41}$$

$$\leq hn^{-1/2}\text{trace}(\boldsymbol{A}_j)\,\|\boldsymbol{\gamma}^*\|_2\,\|\boldsymbol{\Sigma}^*\|_2\,\|\boldsymbol{\theta}^*\|_2 = o(1). \tag{42}$$

Furthermore, we can show that the variance of $b_{j,2}$ is bounded in $N$. Note that $b_{j,1}$ is a quadratic form. Using that $\mathbb{E}[\boldsymbol{V}_j - \boldsymbol{X}_j\boldsymbol{\gamma}^*] = 0$ and $\text{Cov}[\boldsymbol{V}_j - \boldsymbol{X}_j\boldsymbol{\gamma}^*] = \sigma_\epsilon^2\boldsymbol{P}_j^{-1} + o(1)$, it follows that $\mathbb{E}[b_{j,1}] = \text{trace}(\sigma_\epsilon^2\boldsymbol{P}_j^{-1}\boldsymbol{A}_j) + o(1)$ and

$$\text{Var}[b_{j,2}] = \text{trace}(\sigma_\epsilon^2\boldsymbol{P}_j^{-1}\boldsymbol{A}_j\sigma_\epsilon^2\boldsymbol{P}_j^{-1}\boldsymbol{A}_j)$$

which is bounded in $N$. By consequence the mean of the random variable $b_j$ is given by $\mathbb{E}[b_j] = \text{trace}(\sigma_\epsilon^2\boldsymbol{P}_j^{-1}\boldsymbol{A}_j) + o(1)$ and its variance is bounded in $N$. It follows from Markov's law of large numbers that

$$\frac{1}{N}\sum_{j=1}^{N}\left(b_j - \text{trace}(\sigma_\epsilon^2\boldsymbol{P}_j^{-1}\boldsymbol{A}_j)\right) \xrightarrow{p} 0.$$

This implies that $n^{-1}\boldsymbol{V}^\top\boldsymbol{A}(\boldsymbol{V} - \boldsymbol{X}\boldsymbol{\gamma}^*) = n^{-1}\text{trace}(\sigma_\epsilon^2\boldsymbol{P}^{-1}\boldsymbol{A}) + o_P(1)$ which proves (40). This ends the proof of Lemma 5. □

The following lemma is used to derive the properties of the Dantzig-like estimators $\widehat{\boldsymbol{\gamma}}$ and $\widehat{\boldsymbol{\theta}}$.

**Lemma 6.** *Let $\boldsymbol{Y} \in \mathbb{R}^n$, $\boldsymbol{X} \in \mathbb{R}^{n\times p}$ and $C \subseteq \mathbb{R}^p$. Let $\boldsymbol{A} \in \mathbb{R}^{n\times n}$ satisfying the $P$-condition. Suppose there exists $\boldsymbol{\beta}^* \in C$ such that $\left\|n^{-1}\boldsymbol{X}^\top\boldsymbol{A}(\boldsymbol{Y} - \boldsymbol{X}\boldsymbol{\beta}^*)\right\|_\infty \leq \eta$. Let $\widehat{\boldsymbol{\beta}} = \text{argmin}_{\boldsymbol{\beta}\in C}\|\boldsymbol{\beta}\|_1$ subject to $\left\|n^{-1}\boldsymbol{X}^\top\boldsymbol{A}(\boldsymbol{Y} - \boldsymbol{X}\boldsymbol{\beta})\right\|_\infty \leq \eta$. If $s = \|\boldsymbol{\beta}^*\|_0$ and if there exists $\kappa > 0$ such that the restricted eigenvalue condition holds for $(s, \kappa, \boldsymbol{A}^{1/2}\boldsymbol{X})$, this is*

$$\min_{\substack{J_0\subseteq\{1,\ldots,p\}\\|J_0|\leq s}}\ \min_{\substack{\boldsymbol{\delta}\neq 0\\ \left\|\boldsymbol{\delta}_{J_0^c}\right\|_1\leq\|\boldsymbol{\delta}_{J_0}\|_1}} \frac{\left\|\boldsymbol{A}^{1/2}\boldsymbol{X}\boldsymbol{\delta}\right\|_2}{\sqrt{n}\,\|\boldsymbol{\delta}_{J_0}\|_2} \geq \kappa$$

*then for $\boldsymbol{\delta} = \widehat{\boldsymbol{\beta}} - \boldsymbol{\beta}^*$ it holds that $\|\boldsymbol{\delta}\|_1 \leq 8\eta s\kappa^{-2}$ and $n^{-1}\boldsymbol{\delta}^\top\boldsymbol{X}\boldsymbol{A}\boldsymbol{X}\boldsymbol{\delta} \leq 16\eta^2 s\kappa^{-2}$*

**Proof of Lemma 6.** First note that $\boldsymbol{\beta}^*$ satisfies the constraints of the optimization problem whose $\widehat{\boldsymbol{\beta}}$ is the optimal solution. This implies that

$$\left\|\widehat{\boldsymbol{\beta}}\right\|_1 \leq \|\boldsymbol{\beta}^*\|_1. \tag{43}$$

Secondly observe that for the true active set $S_0$ of $\boldsymbol{\beta}^*$ it holds that

$$\left\|\boldsymbol{\beta}^* + \boldsymbol{\delta}_{S_0^c}\right\|_1 = \|\boldsymbol{\beta}^*\|_1 + \left\|\boldsymbol{\delta}_{S_0^c}\right\|_1. \tag{44}$$

Thirdly, combining (43), (44) and the triangle inequality $|a| - |b| \leq |a+b|$ with $a = \boldsymbol{\beta}^* + \boldsymbol{\delta}_{S_0^c}$ and $b = \boldsymbol{\delta}_{S_0}$ we obtain

$$\left\|\boldsymbol{\delta}_{S_0^c}\right\|_1 \leq \|\boldsymbol{\delta}_{S_0}\|_1. \tag{45}$$

Indeed, we have

$$\|\boldsymbol{\beta}^*\|_1 + \left\|\boldsymbol{\delta}_{S_0^c}\right\|_1 - \|\boldsymbol{\delta}_{S_0}\|_1 = \left\|\boldsymbol{\beta}^* + \boldsymbol{\delta}_{S_0^c}\right\|_1 - \|\boldsymbol{\delta}_{S_0}\|_1 \leq \left\|\boldsymbol{\beta}^* + \boldsymbol{\delta}_{S_0^c} + \boldsymbol{\delta}_{S_0}\right\|_1 = \left\|\widehat{\boldsymbol{\beta}}\right\|_1 \leq \|\boldsymbol{\beta}^*\|_1.$$

Now, using the restricted eigenvalue condition together with (45), we have

$$\left\| \boldsymbol{A}^{1/2} \boldsymbol{X} \boldsymbol{\delta} \right\|_2 \geq \sqrt{n}\kappa \left\| \boldsymbol{\delta}_{S_0} \right\|_2 . \tag{46}$$

We can then derive the following result:

$$\begin{aligned} n^{-1} \left\| \boldsymbol{A}^{1/2} \boldsymbol{X} \boldsymbol{\delta} \right\|_2^2 &= n^{-1} \boldsymbol{\delta}^\top \boldsymbol{X}^\top \boldsymbol{A} \boldsymbol{X} \boldsymbol{\delta} \\ &\leq \|\boldsymbol{\delta}\|_1 \left\| n^{-1} \boldsymbol{X}^\top \boldsymbol{A} \boldsymbol{X} (\widehat{\boldsymbol{\beta}} - \boldsymbol{\beta}^*) \right\|_\infty \\ &\overset{(i)}{\leq} 2\eta \|\boldsymbol{\delta}\|_1 \\ &\overset{(ii)}{\leq} 4\eta\sqrt{s} \left\| \boldsymbol{\delta}_{S_0} \right\|_2 \\ &\overset{(iii)}{\leq} 4\eta\sqrt{s}\kappa^{-1} n^{-1/2} \left\| \boldsymbol{A}^{1/2} \boldsymbol{X} \boldsymbol{\delta} \right\|_2 \end{aligned}$$

where (i) follows from the definitioin of $\eta$, (ii) follows from (45) and $\|\boldsymbol{\delta}_{S_0}\|_1 \leq \sqrt{s}\|\boldsymbol{\delta}_{S_0}\|_2$, and (iii) follows from (46). This implies that $n^{-1/2} \left\| \boldsymbol{A}^{1/2} \boldsymbol{X} \boldsymbol{\delta} \right\|_2 \leq 4\eta\sqrt{s}\kappa^{-1}$ and $\|\boldsymbol{\delta}\|_1 \leq 8\eta s\kappa^{-2}$, which ends the proof of Lemma 6. □

**Lemma 7.** *Under conditions of Theorem 4 and for* $\widehat{\sigma}^2 = n^{-1} \left\| \widetilde{\boldsymbol{P}}(\boldsymbol{V} - \boldsymbol{X}\widehat{\gamma}) \right\|_2^2$ *and* $\boldsymbol{K} = \sigma_\epsilon^2 \widetilde{\boldsymbol{P}} \boldsymbol{P}^{-1} \widetilde{\boldsymbol{P}}$, *it holds that* $\widehat{\sigma}^2 = n^{-1}\mathrm{trace}(\boldsymbol{K}) + o_P(1)$.

**Proof of Lemma 7.** Using $\boldsymbol{V} = \boldsymbol{X}\boldsymbol{\gamma}^* + hn^{-1/2}\boldsymbol{Z} + \boldsymbol{W}\boldsymbol{b} + \epsilon$ and writing $\boldsymbol{\epsilon}^* = \widetilde{\boldsymbol{P}}(\boldsymbol{W}\boldsymbol{b} + \boldsymbol{\epsilon})$, we first observe that

$$\begin{aligned} \widehat{\sigma}^2 &= n^{-1} \left\| \widetilde{\boldsymbol{P}}(\boldsymbol{V} - \boldsymbol{X}\widehat{\gamma}) \right\|_2^2 \\ &= n^{-1} \left\| \widetilde{\boldsymbol{P}}(\boldsymbol{W}\boldsymbol{b} + \boldsymbol{\epsilon}) + \widetilde{\boldsymbol{P}}\boldsymbol{X}(\gamma^* - \widehat{\gamma}) + hn^{-1/2}\widetilde{\boldsymbol{P}}\boldsymbol{Z} \right\|_2^2 \\ &= n^{-1} \left\| \widetilde{\boldsymbol{P}}(\boldsymbol{W}\boldsymbol{b} + \boldsymbol{\epsilon}) \right\|_2^2 + o_P(1) \\ &= n^{-1} \boldsymbol{\epsilon}^{*\top} \boldsymbol{\epsilon}^* + o_P(1) \end{aligned} \tag{47}$$

since $n^{-1} \left\| \widetilde{\boldsymbol{P}}\boldsymbol{X}(\gamma^* - \widehat{\gamma}) \right\|_2^2 \leq \lambda_{\max}(\widetilde{\boldsymbol{P}})\, n^{-1} \left\| \widetilde{\boldsymbol{P}}^{1/2}\boldsymbol{X}(\gamma^* - \widehat{\gamma}) \right\|_2^2 = o_P(1)$ by (26) and by the fact that $\lambda_{\max}(\widetilde{\boldsymbol{P}})$ is bounded in $n$ and since $n^{-1/2} \left\| hn^{-1/2}\widetilde{\boldsymbol{P}}\boldsymbol{Z} \right\|_2 \leq hn^{-1} \left\| \widetilde{\boldsymbol{P}} \right\|_2 \|\boldsymbol{Z}\|_2 = o_P(1)$ by the sub-Gaussianity of $z_i$.

Let us define $\boldsymbol{K} = \sigma_\epsilon^2 \widetilde{\boldsymbol{P}} \boldsymbol{P}^{-1} \widetilde{\boldsymbol{P}}$ and observe that $\mathrm{Var}(\boldsymbol{\epsilon}^*) = \boldsymbol{K}$. Working at the group level induced by the mixed model, we consider $\left\{ \boldsymbol{\epsilon}_j^* \right\}_{j=1}^N$ and corresponding covariance matrices $\{\boldsymbol{K}_j\}_{j=1}^N$. We observe that $\left\{ \boldsymbol{\epsilon}_j^{*\top} \boldsymbol{\epsilon}_j^* \right\}_{j=1}^N$ is a sequence of independent random variables with mean $\mu_j = \mathbb{E}[\boldsymbol{\epsilon}_j^{*\top} \boldsymbol{\epsilon}_j^*] = \mathrm{trace}(\boldsymbol{K}_j)$ and $\mathrm{Var}[\boldsymbol{\epsilon}_j^{*\top} \boldsymbol{\epsilon}_j^*] = 2\mathrm{trace}(\boldsymbol{K}_j\boldsymbol{K}_j) < \infty$. The Markov's law of large numbers implies that

$$\frac{1}{N} \sum_{j=1}^N \boldsymbol{\epsilon}_j^{*\top} \boldsymbol{\epsilon}_j^* - \frac{1}{N} \sum_{j=1}^N \mathrm{trace}(\boldsymbol{K}_j) \xrightarrow{p} 0.$$

This together with (47) concludes the proof of Lemma 7. □

**Lemma 8.** *Under conditions of Theorem 4 and for* $\boldsymbol{K} = \sigma_\epsilon^2 \widetilde{\boldsymbol{P}} \boldsymbol{P}^{-1} \widetilde{\boldsymbol{P}}$ *and* $\boldsymbol{w} = n^{-1/2} \widehat{\sigma}_u^{-1} (\boldsymbol{Z} - \boldsymbol{X}\widehat{\boldsymbol{\theta}})$, *it holds that* $\boldsymbol{w}^\top \boldsymbol{K} \boldsymbol{w} = n^{-1} \mathrm{trace}(\boldsymbol{K}) + o_P(1)$.

**Proof of Lemma 8.** Let us define $\boldsymbol{w}^* = n^{-1/2} \sigma_u^{-1} (\boldsymbol{Z} - \boldsymbol{X}\boldsymbol{\theta}^*)$ and observe that $E[\boldsymbol{w}^*] = 0$ and $\mathrm{Var}(\boldsymbol{w}^*) = n^{-1} \boldsymbol{I}_n$ and

$$\boldsymbol{w} = \frac{\sigma_u}{\widehat{\sigma}_u} \left\{ \boldsymbol{w}^* - n^{-1/2} \sigma_u^{-1} \boldsymbol{X} (\widehat{\boldsymbol{\theta}} - \boldsymbol{\theta}^*) \right\}.$$

We consider the following decomposition:

$$\boldsymbol{w}^\top \boldsymbol{K} \boldsymbol{w} = \left( \frac{\sigma_u}{\widehat{\sigma}_u} \right)^2 \Bigg\{ \underbrace{\boldsymbol{w}^{*\top} \boldsymbol{K} \boldsymbol{w}^*}_{T_1} - \underbrace{2 n^{-1/2} \sigma_u^{-1} \boldsymbol{w}^{*\top} \boldsymbol{K} \boldsymbol{X} (\widehat{\boldsymbol{\theta}} - \boldsymbol{\theta}^*)}_{T_2}$$
$$+ \underbrace{n^{-1} \sigma_u^{-2} \left\{ \boldsymbol{X} (\widehat{\boldsymbol{\theta}} - \boldsymbol{\theta}^*) \right\}^\top \boldsymbol{K} \boldsymbol{X} (\widehat{\boldsymbol{\theta}} - \boldsymbol{\theta}^*)}_{T_3} \Bigg\}$$

First, notice that $T_1$ is a quadratic form whose mean and variance are given by:

$$\mathbb{E}[\boldsymbol{w}^{*\top} \boldsymbol{K} \boldsymbol{w}^*] = \mathrm{trace}(n^{-1} \boldsymbol{I}_n \boldsymbol{K}) = n^{-1} \mathrm{trace}(\boldsymbol{K})$$

and

$$\mathrm{Var}[\boldsymbol{w}^{*\top} \boldsymbol{K} \boldsymbol{w}^*] = 2 \mathrm{trace}(n^{-1} \boldsymbol{I}_n \boldsymbol{K} n^{-1} \boldsymbol{I}_n \boldsymbol{K}) = 2 n^{-2} \mathrm{trace}(\boldsymbol{K}\boldsymbol{K}) = O(n^{-1})$$

where the block diagonal structure of $\boldsymbol{K}$ is used to guarantee that the elements of $\boldsymbol{K}\boldsymbol{K}$ are bounded in $n$. Thus the random variable $\boldsymbol{w}^{*\top} \boldsymbol{K} \boldsymbol{w}^* - n^{-1} \mathrm{trace}(\boldsymbol{K})$ has mean 0 and variance converging to 0. It follows from Chebyshev's inequality that

$$T_1 = n^{-1} \mathrm{trace}(\boldsymbol{K}) + o_P(1). \tag{48}$$

Secondly, observe that $T_2$ can be treated as $T_{32}$ in the proof of Theorem 2. Indeed using that $\boldsymbol{w}^* = n^{-1/2} \sigma_u^{-1} \boldsymbol{u}$, we have

$$\begin{aligned} |T_2| &= \left| 2 \sigma_u^{-2} n^{-1} \boldsymbol{u}^\top \boldsymbol{K} \boldsymbol{X} (\widehat{\boldsymbol{\theta}} - \boldsymbol{\theta}^*) \right| \\ &\leq 2 \sigma_u^{-2} n^{-1} \left\| \boldsymbol{K} \boldsymbol{u} \right\|_2 \left\| \boldsymbol{X} (\widehat{\boldsymbol{\theta}} - \boldsymbol{\theta}^*) \right\|_2 = \mathcal{O}_P(\eta_\theta \sqrt{\|\boldsymbol{\theta}^*\|_0}) = o_P(1) \end{aligned} \tag{49}$$

where we use (26) and the sub-Gaussianity of $\boldsymbol{K}\boldsymbol{u}$.

Thirdly, observe that

$$|T_3| = \sigma_u^{-2} n^{-1} \left\| \boldsymbol{K}^{1/2} \boldsymbol{X} (\widehat{\boldsymbol{\theta}} - \boldsymbol{\theta}^*) \right\|_2^2 \leq \sigma_u^{-2} \left\| \boldsymbol{K}^{1/2} \right\|_2^2 n^{-1} \left\| \boldsymbol{X} (\widehat{\boldsymbol{\theta}} - \boldsymbol{\theta}^*) \right\|_2^2 = o_P(1). \tag{50}$$

where the last equality follows from (26) and the fact that $\left\| \boldsymbol{K}^{1/2} \right\|_2^2 \leq \lambda_{\max}(\boldsymbol{K})$ is bounded in $n$.

Since $\widehat{\sigma}_u = \sigma_u + o_P(1)$, the continuous mapping theorem implies that $\frac{\sigma_u}{\widehat{\sigma}_u} \xrightarrow{p} 1$. Hence, this together with (48), (49), (50) and Slutzky's theorem we obtain that

$$\boldsymbol{w}^\top \boldsymbol{K} \boldsymbol{w} = \left( \frac{\sigma_u}{\widehat{\sigma}_u} \right)^2 (T_1 + T_2 + T_3) = n^{-1} \mathrm{trace}(\boldsymbol{K}) + o_P(1). \quad \square$$

The following Lemma was presented in a preliminary version of Zhu and Bradic (2016) and is used in the proof of Theorem 4. It is presented here for the sake of completeness.

**Lemma 9.** *Let $\{w_j\}_{j=1}^n$ and $\{h_j\}_{j=1}^n$ be two independent sequences of random variables. Suppose that*

*(i) $h_j$ and $h_{j'}$ are independent for $j \neq j'$, $\mathbb{E}[h_j] = 0$, $\sigma_h^2 = \mathbb{E}[h_j^2] = \mathcal{O}(1)$ and $\kappa_4 = \mathbb{E}[(h_j^2 - \sigma_h^2)^2] = \mathcal{O}(1)$*

*(ii) $\sum_{j=1}^n w_j^2 = 1$, $\max_{j=1,\dots,n} |w_j h_j| = o_P(1)$ and $\max_{j=1,\dots,n} w_j^2 = o_P(1)$.*

*Then, $\sum_{j=1}^n w_j h_j \xrightarrow{d} \mathcal{N}(0, \sigma_h^2)$.*

The following lemma is an extension of Lemma 9 to the multivariate case. It is useful to obtain the normality of $\widehat{\sigma}_u^{-1} T_1$ in the proof of Theorem 4 where it is used with $\boldsymbol{w} = n^{-1/2}\widehat{\sigma}_u^{-1}(\boldsymbol{Z} - \boldsymbol{X}\widehat{\boldsymbol{\theta}})$ and $\boldsymbol{h} = \widetilde{\boldsymbol{P}}(\boldsymbol{W}\boldsymbol{b} + \boldsymbol{\epsilon})$.

**Lemma 10.** *Let $\{\boldsymbol{w}_j\}_{j=1}^N$ and $\{\boldsymbol{h}_j\}_{j=1}^N$ be two independent sequences of random vectors with $\boldsymbol{w}_j, \boldsymbol{h}_j \in \mathbb{R}^{n_j}$ and $n_j$ bounded in $N$. Suppose that*

*(i) $\boldsymbol{h}_j$ and $\boldsymbol{h}_{j'}$ are independent for $j \neq j'$, $\mathbb{E}[\boldsymbol{h}_j] = 0$, $\mathbb{E}[\boldsymbol{h}_j \boldsymbol{h}_j^\top] = \boldsymbol{K}_j$ with*

$$\lambda_{\max}(\boldsymbol{K}_j) = \mathcal{O}(1)$$

*and*

$$\mathbb{E}[(h_j^i h_j^k - K_j^{ik})(h_j^l h_j^m - K_j^{lm})] = \mathcal{O}(1)$$

*for $1 \leq i, k, l, m \leq n_j$.*

*(ii) $\sum_{j=1}^N \boldsymbol{w}_j^\top \boldsymbol{w}_j = 1$, $\max_{j=1,\dots,N} |\boldsymbol{w}_j^\top \boldsymbol{h}_j| = o_P(1)$ and $\max_{j=1,\dots,N} \boldsymbol{w}_j^\top \boldsymbol{w}_j = o_P(1)$.*

*Then, $\dfrac{\sum_{j=1}^N \boldsymbol{w}_j^\top \boldsymbol{h}_j}{\sqrt{\sum_{j=1}^N \boldsymbol{w}_j^\top \boldsymbol{K}_j \boldsymbol{w}_j}} \xrightarrow{d} \mathcal{N}(0, 1)$.*

**Proof of Lemma 10.** Let $\mathcal{F}_{N,0}$ be the $\sigma$-algebra generated by $W = \{\boldsymbol{w}_j\}_{j=1}^N$ and $\mathcal{F}_{N,j}$ be the $\sigma$-algebra generated by $W$ and $\{\boldsymbol{h}_k\}_{k=1}^j$. Define $X_{Nj} = \boldsymbol{w}_j^\top \boldsymbol{h}_j$ and $S_{Nj} = \sum_{k=1}^j \boldsymbol{w}_k^\top \boldsymbol{h}_k$, $U_{Nj}^2 = \sum_{k=1}^j (\boldsymbol{w}_k^\top \boldsymbol{h}_k)^2$ and $u_N^2 = \sum_{j=1}^N \boldsymbol{w}_j^\top \boldsymbol{K}_j \boldsymbol{w}_j$.

By Theorem 3.4 of Hall and Heyde (1980) it holds that $S_{Nj_N}/U_{Nj_N} \xrightarrow{d} \mathcal{N}(0,1)$ if the following conditions are satisfied: (a) $\max_j |X_{Nj}| = o_P(1)$, (b) $\mathbb{E}\left[\max_j X_{Nj}^2\right]$ is bounded in $N$, (c) $U_{Nj_N}^2 - u_N^2 = o_P(1)$, (d) $\sum_{j=1}^N \mathbb{E}\left[X_{Nj} | \mathcal{F}_{N,j-1}\right] = o_P(1)$ and (e) $\sum_{j=1}^N \left|\mathbb{E}\left[X_{Nj} | \mathcal{F}_{N,j-1}\right]\right|^2 = o_P(1)$.

Observe that (a) holds from the assumptions of the Lemma. Secondly, notice that

$$\mathbb{E}\left[\max_j (\boldsymbol{w}_j^\top \boldsymbol{h}_j)^2\right] \leq \sum_{j=1}^N \mathbb{E}\left[(\boldsymbol{w}_j^\top \boldsymbol{h}_j)^2\right] \tag{51}$$

$$= \sum_{j=1}^N \mathbb{E}\left[\mathbb{E}\left[(\boldsymbol{w}_j^\top \boldsymbol{h}_j)^2 | \mathcal{F}_{N,j-1}\right]\right] = \sum_{j=1}^N \mathbb{E}\left[\boldsymbol{w}_j^\top \boldsymbol{K}_j \boldsymbol{w}_j\right]. \tag{52}$$

Since $\boldsymbol{w}_j^\top \boldsymbol{K}_j \boldsymbol{w}_j = \left\|\boldsymbol{K}_j^{1/2}\boldsymbol{w}_j\right\|_2^2$, we derive that

$$\sum_{j=1}^{N} \mathbb{E}\left[\boldsymbol{w}_j^\top \boldsymbol{K}_j \boldsymbol{w}_j\right] \leq \sum_{j=1}^{N} \mathbb{E}\left[\left\|\boldsymbol{K}_j^{1/2}\right\|_2^2 \|\boldsymbol{w}_j\|_2^2\right] \tag{53}$$

$$\leq \max_{j=1,\ldots,N} \lambda_{\max}(\boldsymbol{K}_j)\, \mathbb{E}\left[\sum_{j=1}^{N} \boldsymbol{w}_j^\top \boldsymbol{w}_j\right] < \infty, \tag{54}$$

where we use $\sum_{j=1}^N \boldsymbol{w}_j^\top \boldsymbol{w}_j = 1$ and the fact that $\max_j \lambda_{\max}(\boldsymbol{K}_j)$ is bounded in $N$. Thus, (b) is proven. To show (c), we define $Q_N = \sum_{j=1}^N \boldsymbol{w}_j^\top(\boldsymbol{h}_j\boldsymbol{h}_j^\top - \boldsymbol{K}_j)\boldsymbol{w}_j$ and need to show that $Q_N = o_P(1)$. First, we derive that

$$\begin{aligned}
\mathbb{E}[Q_N^2|W] &= \sum_{j=1}^{N}\sum_{k=1}^{N} \mathbb{E}[\boldsymbol{w}_j^\top(\boldsymbol{h}_j\boldsymbol{h}_j^\top - \boldsymbol{K}_j)\boldsymbol{w}_j\boldsymbol{w}_k^\top(\boldsymbol{h}_k\boldsymbol{h}_k^\top - \boldsymbol{K}_k)\boldsymbol{w}_k|W] \\
&\overset{(i)}{=} \sum_{j=1}^{N} \mathbb{E}[\boldsymbol{w}_j^\top(\boldsymbol{h}_j\boldsymbol{h}_j^\top - \boldsymbol{K}_j)\boldsymbol{w}_j\boldsymbol{w}_j^\top(\boldsymbol{h}_j\boldsymbol{h}_j^\top - \boldsymbol{K}_j)\boldsymbol{w}_j|W] \\
&\overset{(ii)}{=} \sum_{j=1}^{N} \operatorname{trace}\left(\boldsymbol{w}_j\boldsymbol{w}_j^\top \operatorname{Cov}[(\boldsymbol{h}_j\boldsymbol{h}_j^\top - \boldsymbol{K}_j)\boldsymbol{w}_j|W]\right)
\end{aligned}$$

where $(i)$ follows from the independence assumption of $\boldsymbol{h}_j$ and $(ii)$ follows from the mean formula of a quadratic form and from $\mathbb{E}[(\boldsymbol{h}_j\boldsymbol{h}_j^\top - \boldsymbol{K}_j)\boldsymbol{w}_j|W] = 0$. Let us now define the $n_j^2 \times n_j$ matrix

$$\boldsymbol{w}_j^{\text{diag}} = \begin{pmatrix} \boldsymbol{w}_j & \boldsymbol{0}_{n_j} & \ldots & \boldsymbol{0}_{n_j} \\ \boldsymbol{0}_{n_j} & \boldsymbol{w}_j & \ldots & \boldsymbol{0}_{n_j} \\ \vdots & \vdots & \vdots & \vdots \\ \boldsymbol{0}_{n_j} & \boldsymbol{0}_{n_j} & \ldots & \boldsymbol{w}_j \end{pmatrix}$$

and notice that

$$\operatorname{Cov}[(\boldsymbol{h}_j\boldsymbol{h}_j^\top - \boldsymbol{K}_j)\boldsymbol{w}_j|W] = \boldsymbol{w}_j^{\text{diag}\top} \mathcal{K}_{4,j}\, \boldsymbol{w}_j^{\text{diag}}$$

where $\mathcal{K}_{4,j}$ is a $n_j^2 \times n_j^2$ matrix containing all the fourth moment elements of $\boldsymbol{h}_j$. By assumption (i), all the elements of $\mathcal{K}_{4,j}$ are bounded. It follows that

$$\begin{aligned}
\mathbb{E}[Q_N^2|W] &= \sum_{j=1}^{N} \boldsymbol{w}_j^\top \boldsymbol{w}_j^{\text{diag}\top} \mathcal{K}_{4,j}\, \boldsymbol{w}_j^{\text{diag}} \boldsymbol{w}_j \\
&= \sum_{j=1}^{N} \left\|\mathcal{K}_{4,j}^{1/2} \boldsymbol{w}_j^{\text{diag}} \boldsymbol{w}_j\right\|_2^2 \\
&\leq \sum_{j=1}^{N} \left\|\mathcal{K}_{4,j}^{1/2}\right\|_2^2 \left\|\boldsymbol{w}_j^{\text{diag}}\right\|_2^2 \|\boldsymbol{w}_j\|_2^2 \\
&\leq \sum_{j=1}^{N} C\, n_j^2\, (\boldsymbol{w}_j^\top \boldsymbol{w}_j)^2
\end{aligned}$$

$$\leq C\, n_j^2 \max_{j=1,\ldots,N} (\boldsymbol{w}_j^\top \boldsymbol{w}_j) \sum_{j=1}^N \boldsymbol{w}_j^\top \boldsymbol{w}_j$$

$$= o_P(1),$$

for a constant $C$, where we use the cycling property of the trace, the boundedness of the fourth moments of $\boldsymbol{h}_j$, the property $\sum_j c_j^2 \leq \max |c_j| \ \sum_j |c_j|$ and the assumption (ii). By consequent, we have

$$\mathbb{E}[Q_N^2] = \mathbb{E}[\mathbb{E}[Q_N^2 | W]] = o_P(1)$$

which implies by Markov's inequality that $Q_N^2 = o_P(1)$. This proves (c).

Regarding (d) and (e), we observe that

$$\mathbb{E}\left[X_{Nj} | \mathcal{F}_{N,j-1}\right] = \mathbb{E}\left[\boldsymbol{w}_j^\top \boldsymbol{h}_j | \mathcal{F}_{N,j-1}\right] = \boldsymbol{w}_j^\top \mathbb{E}\left[\boldsymbol{h}_j | \mathcal{F}_{N,j-1}\right] = 0$$

by the independence assumption between $\boldsymbol{h}_j$ and $\boldsymbol{h}_1, \ldots, \boldsymbol{h}_{j-1}$ and the independence between $\boldsymbol{h}_j$ and $\boldsymbol{w}_j$.

This guarantees that (d) and (e) hold. We thus obtain that $S_{Nj_N}/U_{Nj_N} \xrightarrow{d} \mathcal{N}(0,1)$. This together with (c) and Slutsky's Theorem implies that

$$\frac{\sum_{j=1}^N \boldsymbol{w}_j^\top \boldsymbol{h}_j}{\sqrt{\sum_{j=1}^N \boldsymbol{w}_j^\top \boldsymbol{K}_j \boldsymbol{w}_j}} \xrightarrow{d} \mathcal{N}(0,1).$$

□